\documentclass[pdflatex,sn-basic]{sn-jnl}
\usepackage{amsmath,amssymb}
\usepackage{graphicx}
\usepackage{bm}
\usepackage{xcolor}
\usepackage{booktabs}
\usepackage{siunitx}
\usepackage{placeins}
\newcommand{\ReD}{Re_D}
\newcommand{\Ub}{U_b}
\newcommand{\Dpipe}{D}
\newcommand{\Rc}{R_c}
\newcommand{\curv}{\gamma}
\newcommand{\Us}{U_s}
\newcommand{\Ur}{U_r}
\newcommand{\se}{s_{\mathrm{e}}}
\newcommand{\dsd}{\Delta s_{\mathrm{d}}}
\newcommand{\dd}{\mathrm{d}}
\newcommand{\vect}[1]{\mathbf{#1}}

\providecommand{\St}{\mathit{St}}

\begin{document}

\title{Experimental study of the Reynolds number effect on the mean flow and coherent structures in a \SI{180}{\degree} bent pipe flow using PIV}

\author*[1]{\fnm{Eman} \sur{Bagheri}}\email{eman.m.bagheri@fau.de}
\author[1]{\fnm{Christoph} \sur{Heigl}}
\author[1]{\fnm{Felix} \sur{Czwielong}}
\author[1]{\fnm{Stefan} \sur{Becker}}
\author[1]{\fnm{Philipp} \sur{Schlatter}}

\affil*[1]{\orgdiv{Institute of Fluid Mechanics (LSTM)}, \orgname{Friedrich--Alexander--Universität (FAU) Erlangen--Nürnberg}, \orgaddress{\postcode{91058} \city{Erlangen}, \country{Germany}}}

\abstract{\unboldmath In this work, we study the effect of Reynolds number on the flow in a \SI{180}{\degree} pipe bend with a curvature of $\gamma=0.2$. We perform planar particle image velocimetry (PIV) in the symmetry plane of the pipe for bulk Reynolds numbers $\ReD$ from $10\,000$ to $60\,000$ with water as the working fluid. To remove the refraction at the pipe wall, a hybrid method is employed. A mixture of water, glycerol and sodium iodide is used to match the refractive index of the pipe at its outer interface. The refraction at the inner interface is corrected using reverse ray tracing. The measurements are validated against direct numerical simulations at $\ReD=10\,000$ and $20\,000$ and against hot-wire data from the literature at $60\,000$. As the Reynolds number increases, the mean velocity profiles in and downstream of the bend become flatter. Furthermore, a second vortex pair, with the opposite direction of rotation to the Dean vortices, forms further downstream, between bend angles of \SI{116}{\degree} and \SI{126}{\degree}. A modal decomposition of the planar velocity fields shows that the coherent structures keep their spatial features at all Reynolds numbers. Their Strouhal numbers remain nearly unchanged and show no monotonic trend.}


\maketitle

\section{Introduction}
\label{sec:intro}

Flows in curved pipes occur in a wide variety of technical and physiological systems, and have been the subject of research for more than a century \citep{Berger1983,Ito1987} with the seminal work of Dean \citep{Dean1927,Dean1928}. Curved-pipe flows are characterised by two independent non-dimensional numbers. These are the bulk Reynolds number $\ReD=\Ub\Dpipe/\nu$ and the curvature $\curv=\Dpipe/(2\Rc)$. Here, $\Ub$ and $\Dpipe$ denote the bulk velocity and the pipe diameter, respectively. The symbols $\nu$ and $\Rc$ denote the kinematic viscosity and the radius of curvature of the centreline, respectively. In curved pipe flows, the centrifugal force generates a secondary flow in the form of a counter-rotating pair of Dean vortices \citep{Dean1927,Dean1928}. Further downstream in the bend, the cross-stream flow on the symmetry plane reverses its direction. The reversal is attributed to a second pair of counter-rotating vortices with the opposite direction of rotation to the Dean vortices \citep{Rowe1970,Azzola1986,Anwer1989,Lee2007}. In earlier measurements and simulations \citep{Anwer1989,Lai1991,Moulinec2005,Lee2007}, this pair was found to form between bend angles of \SI{67.5}{\degree} and \SI{135}{\degree}. In bends with curvatures of $0.25$ and $0.33$ compared to $0.2$ in this study, the pair appears downstream of the bend exit \citep{Sudo2000,Lupi2025}.

The secondary motion in a bent pipe flow is typically not steady. \citet{Tunstall1968} observed in mitred \SI{90}{\degree} bends that the Dean pair was replaced by a single vortex that intermittently flipped its direction of rotation, at frequencies $f$ corresponding to Strouhal numbers $\St=f\Dpipe/\Ub$ of $2\times10^{-4}$ to $4.5\times10^{-3}$. \citet{Brucker1998} measured an alternating switching of the vortex pair $1.5\Dpipe$ downstream of a \SI{90}{\degree} bend at $\St\approx0.03$ and $0.12$, and named it swirl switching. Several unsteady structures were identified in subsequent studies using time-resolved particle image velocimetry (PIV) \citep{Kalpakli2013,Kalpakli2013_2,Hellstrom2013,KalpakliVester2015}, large-eddy simulation \citep{Rutten2005,Carlsson2015,He2021} and direct numerical simulation (DNS) \citep{Noorani2016_Swirl,Hufnagel2018,Wang2018,Lupi2025}. In some of these works, the term swirl switching was used to refer to coherent structures that differ from those described by \citet{Brucker1998} in frequency, in their spatial features and in their location with respect to the bend. Later, \citet{bagheri2026swirl} separated these structures into four families of distinct modes depending on their location, symmetry, and generation mechanism. They found a pair of swirl-switching and swirl-breathing modes inside the bend and showed that these modes arise from the local instability of the mean flow. Additionally, two shear-layer modes and a low-frequency axial wave were identified at the downstream straight section of the bend.

Most of these experiments were performed at Reynolds numbers of a few times $10^4$ or below, and the direct numerical simulations are limited to lower Reynolds numbers due to their computational cost. Currently, the highest Reynolds number explored in bent pipes using DNS is $\ReD=2\times10^4$ \citep{bagheri2026}, to the best of our knowledge. Velocity data over a wide range of Reynolds numbers exist mainly for \SI{90}{\degree} bends \citep{Hellstrom2013,Haroual2025}, while \SI{180}{\degree} bends are more limited \citep{Rowe1970,Azzola1986,Sudo2000,Jain2022}. As a result, the effect of Reynolds number on the flow of a \SI{180}{\degree} bend has to be inferred from studies at different bend angles and different curvatures. In particular, the effect of Reynolds number on the mean velocity profile remains largely unexplored. This has important practical implications, as flow meters and branch connections are calibrated and designed for fully developed pipe flow, and their intended performance is affected by distorted profiles and swirling flows. Consequently, this can lead to a higher measurement uncertainty or increased pressure losses. This motivates the current work in which the mean flow of a \SI{180}{\degree} bend is measured by planar PIV between $\ReD=10\,000$ and $60\,000$ to study the effect of Reynolds number.

For the PIV measurement of the flow through a pipe, the pipe must be made of an optically transparent material. However, the images can become distorted as the wall acts as a lens with a varying power depending on the local curvature of the wall and the angle at which the light crosses it. As a result, a regular grid in the symmetry plane appears distorted on the camera sensor. Using a calibration target \citep{SoloffAdrianLiu1997Distortion} placed inside the fluid, a mapping can be obtained for image corrections. However, in many applications, placing a calibration target is infeasible. Furthermore, self-calibration using the particle images requires two cameras \citep{Wieneke2005SelfCalibration}. Even when a calibration target is placed inside a straight glass pipe, the mapping deviated by up to \SI{0.27}{mm} in the near-wall region \citep{VanDoorne2007}. As a result, most PIV studies of bent pipe flows were performed on the flow exiting the bend. Several of these studies state that optical access or technical restrictions were the reason for their choice \citep{Kalpakli2013_2,KalpakliVester2015,KalpakliSattarzadeh2016,Sundstrom2026}. Even when the outer refraction was reduced or removed, the measurements close to the wall remained inaccessible due to residual refractions \citep{Hellstrom2013,Sundstrom2026}.

The refraction can largely be removed by matching the refractive index of the liquid to that of the pipe wall \citep{Budwig1994IndexMatching,Wiederseiner2011}. For sodium-iodide solutions, \citet{Bai2014} extended the refractive index up to $1.51$ at the solubility limit at \SI{40}{\celsius}. Index of refraction matching enables planar, high-speed, and tomographic PIV in curved and helically coiled pipes \citep{vanWyk2015,Wang2023,Muller2026b,Muller2026a,Kovats2020}. However, the use of these liquids comes at the price of viscosity, cost, toxicity, or strong temperature sensitivity \citep{Budwig1994IndexMatching,Wiederseiner2011,Wright2017}. As an alternative to physically matching the indices, the refraction can be modelled, and the image correction can be performed numerically. Various camera models are available for stereoscopic PIV, photogrammetry, and tomographic PIV, in which rays are traced through a planar window \citep{Wieneke2005SelfCalibration,Wang2026} or a cylindrical wall \citep{PaolilloAstarita2022Refraction}. \citet{Velten2024RTPIV} performed ray tracing through the glass spheres of a packed bed. They found that the model was highly sensitive to its geometric parameters. Before computing the correlations, \citet{Martins2018_Ray_Tracing_Based}, \citet{Zha2016Distortion} and \citet{Akkurt2022Dimpled} corrected the PIV images using ray tracing, and \citet{Kang2004Droplet} corrected the velocity field inside a droplet. \citet{Paolillo2021_On_the_PIV} showed that high orders are needed when polynomial camera models are used for strongly curved interfaces. For a refractive camera model calibrated over a restricted region, \citet{PaolilloAstarita2022Refraction} showed that extrapolation outside this region is poor. These findings emphasise why the optical geometry needs to be known independently.

In this study, we have an external bath outside the curved pipe containing a liquid with a refractive index matching that of the pipe's material. This physically removes refraction at the outer interface (section~\ref{sec:rim}). However, the working fluid inside the pipe is water, therefore, the refraction at the inner interface remains. We correct this refraction numerically. We fit a reverse ray-tracing model of the computer-aided design (CAD) surfaces to the wall contours detected in the images. In the next step, we invert the fitted model to obtain a resampling map. This map is used to correct every image before performing the correlation computations (section~\ref{sec:raytracing}).

We first validate the planar PIV measurement before drawing any physical conclusions from the results. The measurement is compared with DNS of the same geometry at $\ReD=10\,000$ and $20\,000$. Furthermore, we compare the measurement with the hot-wire data of \citet{Sudo2000} at $60\,000$. In the next step, we investigate how the Reynolds number up to $\ReD=60\,000$ affects the flow. Increasing the Reynolds number beyond $\ReD=20\,000$ becomes prohibitively expensive in the DNS. Therefore, we extend the investigation to higher Reynolds numbers using PIV.

The paper is structured in the following way. The description of the experimental setup is presented in section~\ref{sec:setup}, followed by the ray-tracing and image corrections in section~\ref{sec:raytracing}. Sections~\ref{sec:piv} and~\ref{sec:dns} explain the PIV evaluation and the simulations, respectively. We validate the measurement in section~\ref{sec:validation} and present the results in section~\ref{sec:results}. The results section is divided into two parts. In the first part (section~\ref{sec:sec_flow}), we investigate the effect of Reynolds number on the mean flow and on the onset of the second vortex pair. The second part (section~\ref{sec:fhpod}) discusses the coherent structures identified by \citet{bagheri2026swirl} at $\ReD=10\,000$, where we investigate whether the spatial features and Strouhal numbers of these structures persist up to $\ReD=60\,000$. Finally, the conclusions are given in section~\ref{sec:conclusions}.
\section{Experimental facility and test section}
\label{sec:setup}

\subsection{Experimental facility}

\begin{figure}[tbp]
  \centering
  \begin{minipage}[b]{0.58\linewidth}
  \centering
  \includegraphics{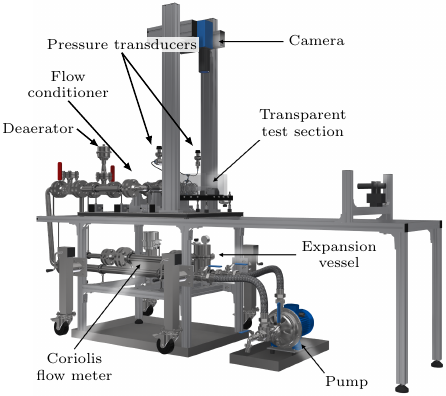}\\[-1pt]
  {\small (a)}
  \end{minipage}\hfill
  \begin{minipage}[b]{0.385\linewidth}
  \centering
  \includegraphics[width=\linewidth]{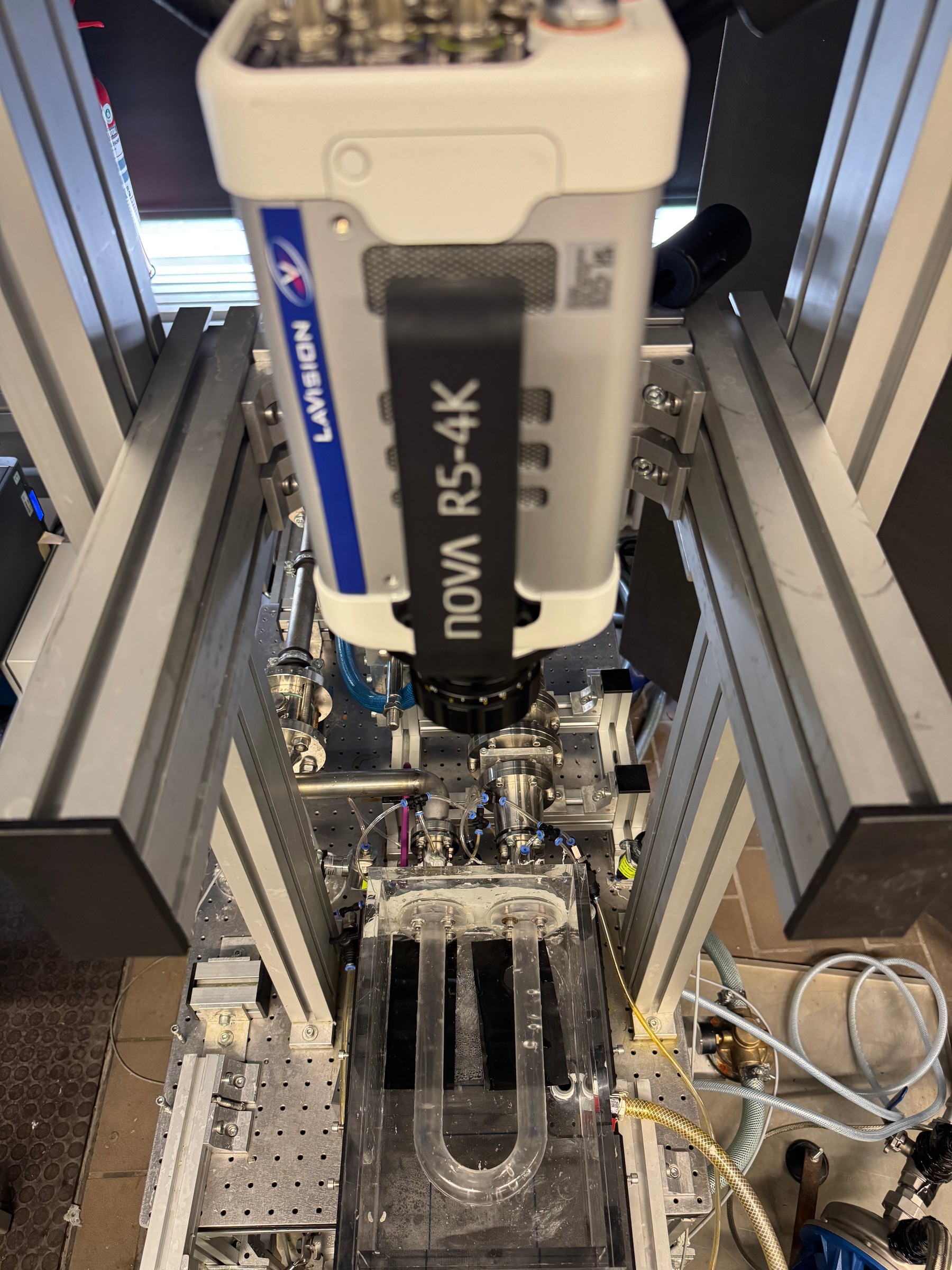}\\[-1pt]
  {\small (b)}
  \end{minipage}
  \caption{Bent pipe test bench: (a) hydraulic and optical components and (b) the camera mounted above the transparent test section.}
  \label{fig:rig}
\end{figure}

The measurements are performed using the bent pipe test bench at the Institute of Fluid Mechanics (LSTM) shown in figure~\ref{fig:rig}(a). The water is pumped in a closed loop using a centrifugal pump, and a membrane expansion vessel is used to attenuate the pressure pulsations caused by the pump. The mass flow rate $\dot m$ is measured with a Coriolis flow meter (type Proline Promass E100, producer Endress+Hauser GmbH+Co. KG), and the entrained air is removed using a deaeration unit. The Coriolis meter has a nominal diameter of \SI{40}{mm}, a measurement range of 0 to \SI{45000}{kg/h}, and a measurement accuracy of $0.1\,\%$ of the measured value. Upstream of the test section, a conditioning section is used to ensure the inflow profile is fully developed at the inlet of the test section. This section trips the flow with a ring that blocks about $30\,\%$ of the cross-section and lets it develop over $80\Dpipe$ of straight pipe \citep{Baumeister2018_MA}, beyond the trip blockage of $15\,\%$ and the \num{60} channel heights that \citet{Durst1998} found sufficient for a fully developed turbulent flow independent of the upstream conditions. A differential pressure sensor (type PD-33X, producer KELLER Pressure AG) is placed between the inlet and outlet planes of the test section, with an absolute pressure sensor (type PAA-33X, producer KELLER Pressure AG) placed at the inlet. The absolute pressure sensor has a measurement range of 0 to \SI{10}{bar} and a measurement accuracy of $0.1\,\%$ FS (full scale). The differential pressure sensor has a differential pressure range of \SI{1}{bar} and a measurement accuracy of $0.1\,\%$ FS. The maximum sampling rate for both pressure sensors is \SI{400}{Hz}. The water temperature is measured using a PT-100 temperature sensor. The mass flow rate and the Reynolds number are controlled using the rotational speed of the pump. The Reynolds number is obtained from the mass flow rate as
\begin{equation}
  \ReD=\frac{\Ub\Dpipe}{\nu}=\frac{4\dot m}{\pi\Dpipe\mu},
  \label{eq:re_massflow}
\end{equation}
in which $\mu$ and $\nu$ denote the dynamic and the kinematic viscosity of water at the measured temperature, respectively. Besides pressure loss measurement, the test rig allows optical measurements using a transparent test section. Figure~\ref{fig:rig}(b) shows the camera mounted above the test section, normal to the laser sheet, enabling PIV measurements on the pipe's symmetry plane.

\subsection{Test section}

\begin{figure}[tbp]
  \centering
  \includegraphics[width=0.95\linewidth]{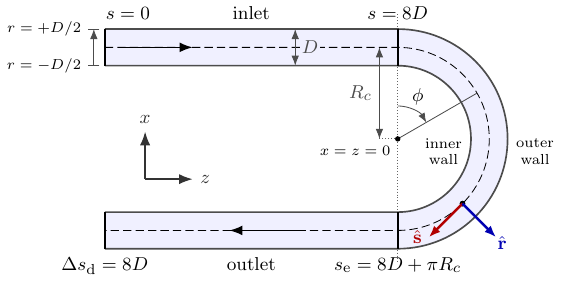}
  \caption{Bend geometry and coordinates in the symmetry plane, with the flow direction (arrows) and the unit vectors $\hat{\vect s}$ and $\hat{\vect r}$ at $\phi=\SI{135}{\degree}$, with the outlet section drawn up to $\dsd=8\Dpipe$.}
  \label{fig:coords}
\end{figure}

The test section is a \SI{180}{\degree} bend with an inner diameter of $\Dpipe=\SI{20}{mm}$ and a centreline radius of $\Rc=\SI{50}{mm}$. The curvature of the bend, $\curv=\Dpipe/(2\Rc)=0.2$, is that of the reference simulations. The straight sections of the transparent test section are \num{8}$\Dpipe$ long and are connected to the conditioning section and to the return line. A schematic of the test section and the coordinate system is shown in figure~\ref{fig:coords}. The streamwise coordinate $s$ follows the pipe centreline along the unit vector $\hat{\vect s}$, and $s=0$ lies \num{8}$\Dpipe$ upstream of the bend entry. The bend is located between $s=8\Dpipe$ and $s=\se$, where $\se=8\Dpipe+\pi\Rc$ is the bend exit. The bend entry is located at $\phi=\SI{0}{\degree}$, and the bend exit lies at $\phi=\SI{180}{\degree}$. The downstream distance from the bend exit is measured as $\dsd=s-\se$.

The laser sheet lies in the symmetry plane of the pipe, which contains the pipe centreline. The radial unit vector $\hat{\vect r}$ is normal to the centreline in the symmetry plane and points towards the outer wall. The time-averaged streamwise and radial velocity components $\Us$ and $\Ur$ are defined along $\hat{\vect s}$ and $\hat{\vect r}$, respectively. This means that a positive $\Ur$ denotes flow towards the outer wall, and a negative $\Ur$ denotes flow towards the inner wall. Along $\hat{\vect r}$, the signed coordinate $r$ is measured from the centreline, with $r=-\Dpipe/2$ at the inner wall and $r=+\Dpipe/2$ at the outer wall. For the contour plots, we use the Cartesian coordinates $z$ and $x$ in the symmetry plane, with their origin at the centre of the bend. Here, $z$ is parallel to the straight sections towards the bend, and $x$ points towards the inlet section.

\begin{figure}[tbp]
  \centering
  \includegraphics[width=0.49\linewidth]{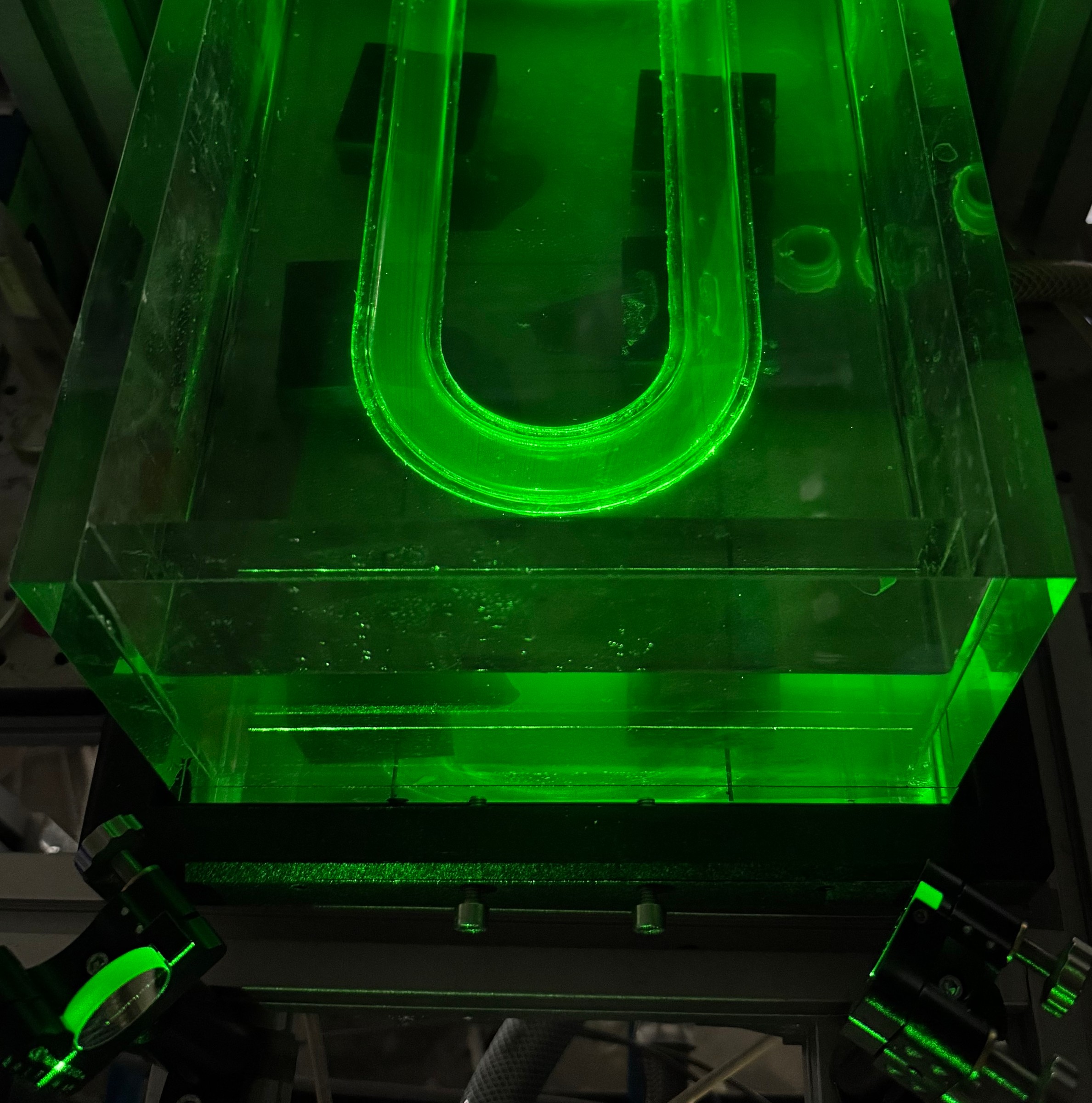}\hspace{0.02\linewidth}\includegraphics[width=0.397\linewidth]{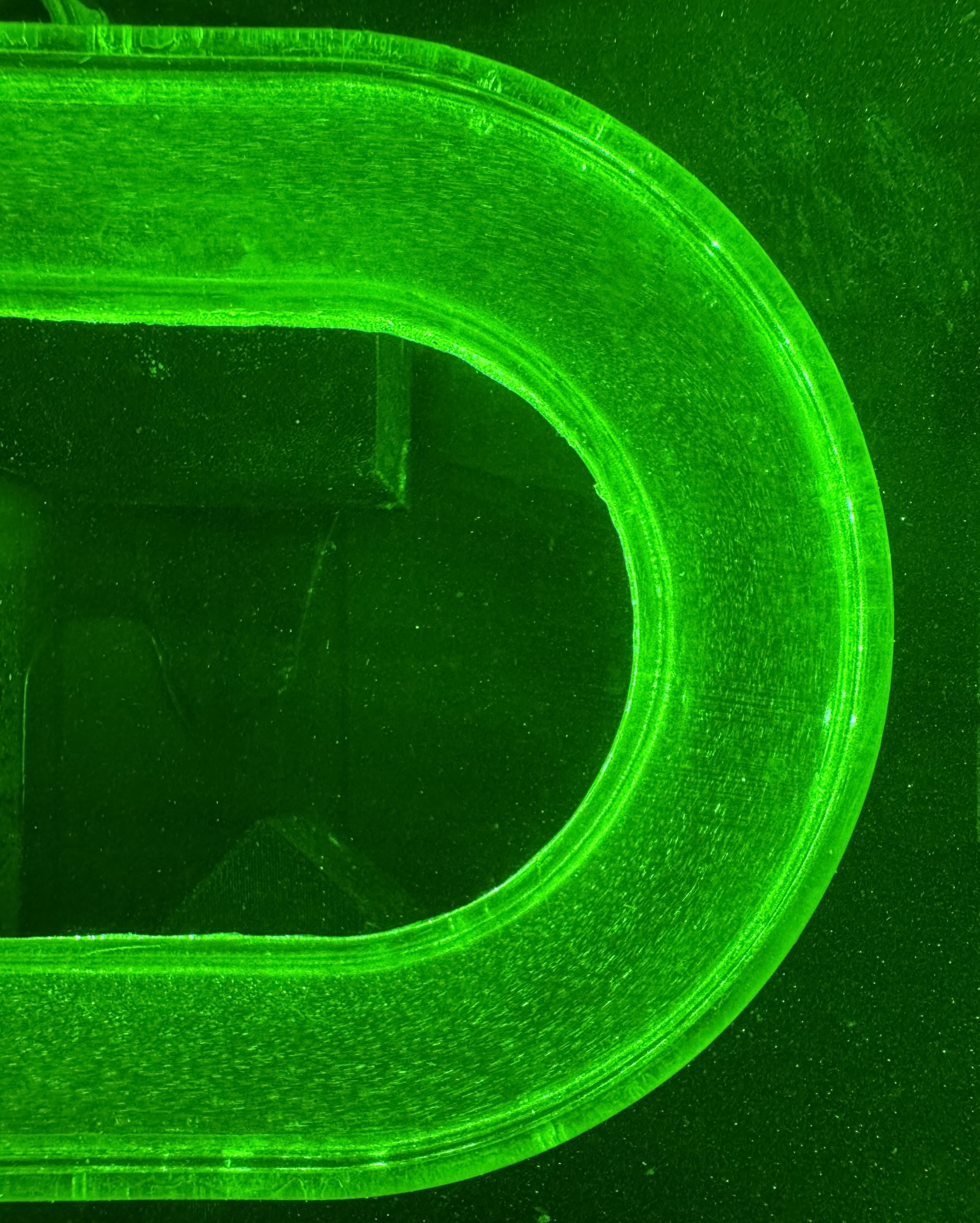}\\[-1pt]
  {\small\makebox[0.49\linewidth]{(a)}\hspace{0.02\linewidth}\makebox[0.397\linewidth]{(b)}}
  \caption{Test section in the laser light: (a) the bend in its bath and (b) the bend seen from above.}
  \label{fig:section}
\end{figure}

The transparent $\SI{180}{\degree}$ pipe bend (figure~\ref{fig:section}(a, b)) is vacuum-cast from a water-clear polyurethane resin with a refractive index of $1.51$ at \SI{20}{\celsius} \citep{SynthenePRC1810}. The cast part retains the CAD geometry within the tolerance of the mould. Therefore, we use the stereolithography (STL) surface model of the CAD geometry to define the optical interfaces for the ray-tracing correction in section~\ref{sec:raytracing}.

\subsection{Refractive-index matching}
\label{sec:rim}

A common solution to the refraction at the curved pipe wall is to match the index of refraction (IOR) of the working fluid to that of the wall \citep{Budwig1994IndexMatching}. However, IOR-matched liquids are in general more viscous than water. For instance, the index of refraction of an aqueous glycerol solution rises from $1.33$ to $1.47$ with the glycerol content, while its viscosity rises to up to $1490$ times that of water \citep{Budwig1994IndexMatching}. The IOR-matched liquid used for this facility is a mixture of sodium iodide, glycerol and water with about twelve times the kinematic viscosity of water \citep{Baumeister2018_MA}. The velocity needed to attain a given Reynolds number is proportional to the kinematic viscosity when the diameter is fixed. As a result, a test rig running on the IOR-matched liquid would have to circulate twelve times the flow rate. As the pressure drop increases with the square of the velocity, the required pumping power would increase by a factor of roughly $12^{3}$. For this reason, we do not use the IOR-matched liquid as the working fluid in the test rig. Instead, we use a hybrid approach in which the refractions at the outer interface are removed using an IOR-matched fluid bath, while the refractions at the inner interface are corrected numerically.

The IOR-matched liquid is kept in the quiescent bath to minimise thermal effects from the pump that could affect the refractive index match \citep{Wiederseiner2011,Wright2017}. The bath consists of \num{53}\,\% sodium iodide, \num{26}\,\% glycerol and \num{21}\,\% water by mass. Mixtures of these components were used for index matching by \citet{Baldwin1989}, \citet{Yousif2011}, \citet{Le2013} and, in a \SI{180}{\degree} bend, by \citet{vanWyk2015}. At \SI{23}{\celsius}, an aqueous sodium-iodide solution reaches an index of $1.499$ at its solubility limit \citep{Bai2014}. Therefore, part of the water is replaced by glycerol, whose refractive index of about $1.47$ exceeds that of water, to raise the index of the mixture above this value. Furthermore, a small amount of sodium thiosulfate keeps the solution colourless \citep{ParkerMerati1996,Narrow2000,Wright2017}.

\section{Reverse ray tracing}
\label{sec:raytracing}

\begin{figure*}[tbp]
  \centering
  \includegraphics{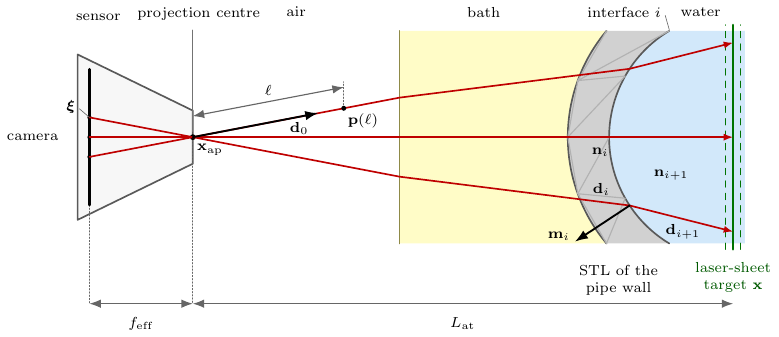}
  \caption{Schematic of the optical model, with the IOR-matched bath in yellow and the water inside the pipe in blue.}
  \label{fig:rt_schematic}
\end{figure*}

Figure~\ref{fig:rt_schematic} shows a schematic sketch of the refractions occurring as a result of the IOR mismatch between different media at the free surface of the bath and at the interface between the pipe and the water inside. Due to the refractions, the recorded images are distorted, and the magnification factor (pixels per millimetre) in the laser sheet is not constant. Cross-correlation of the raw images would therefore measure displacements in this warped space, which differ from the true distances between the particles in an image pair. To account for the distortion, we perform reverse ray tracing and find the local distortion in the image space. The mapping function found using reverse ray tracing is then used to correct the distortions and restore the true image before doing the PIV evaluations. An alternative approach is to correct the velocity vectors afterwards, but this method is limited due to its underlying assumption of a constant magnification over an interrogation window \citep{SoloffAdrianLiu1997Distortion}. However, this assumption does not hold, especially near the walls, where the magnification factor changes within an interrogation window.

\FloatBarrier

\subsection{Ray model}
\label{sec:raymodel}

We use the pinhole model for the camera, in which every ray passes through the projection centre point, at the distance $L_{\mathrm{at}}$ from the laser sheet (figure~\ref{fig:rt_schematic}). The effective focal length $f_{\mathrm{eff}}$ of the pinhole camera is found from separate images of a printed pattern of known spacing, recorded at several measured distances. The effective focal length exceeds the nominal focal length by about $8\,\%$, as the lens is focused at a finite distance \citep{Wieneke2005SelfCalibration,Velten2024RTPIV}.

The sensor is located at the distance $f_{\mathrm{eff}}$ behind the projection centre $\vect x_{\mathrm{ap}}$. The ray of the pixel at the sensor position $\bm\xi$ leaves the projection centre in the unit direction $\vect d_0$ and runs along the points $\vect p(\ell)$,
\begin{equation}
  \vect d_0=\frac{\vect x_{\mathrm{ap}}-\bm\xi}{\lVert\vect x_{\mathrm{ap}}-\bm\xi\rVert},
  \qquad
  \vect p(\ell)=\vect x_{\mathrm{ap}}+\ell\,\vect d_0 ,
  \label{eq:pinhole}
\end{equation}
in which $\ell\ge0$ is the path length from the projection centre to the point $\vect p(\ell)$.

Using the pinhole model, we follow the individual rays in reverse direction from a sensor pixel through the projection centre and the refracting interfaces towards the light sheet. As this direction is opposite to that of the light reaching the camera sensor, the method is referred to as reverse ray tracing. As all rays must pass through the projection centre, every pixel on the image sensor requires a single ray, and the path follows from Snell's law at each interface. In contrast, direct ray tracing from a point of the light sheet to the camera requires an iterative search for the direction that passes through the projection centre. To increase the accuracy of the ray tracing and account for the variation of the surface normal at each triangulated surface panel, we use barycentric interpolation instead of using a single normal per facet used in earlier refractive models \citep{PaolilloAstarita2022Refraction}.

Each ray passes through multiple interfaces before reaching the plane of the laser sheet. We obtain all material interfaces from the triangulated surfaces of the CAD model. At the intersection point of the ray and interface $i$, the unit normal $\vect m_i$ is interpolated between the vertex normals of the triangle. This normal points in the opposite direction to the incident direction $\vect d_i$. Let $\eta_i$ and $\zeta_i$ denote the refractive index ratio and the cosine of the incidence angle, respectively. The transmitted direction can then be obtained from Snell's law
\begin{equation}
  \vect d_{i+1}=\eta_i\vect d_i+\Big(\eta_i\zeta_i-\sqrt{1-\eta_i^2\big(1-\zeta_i^2\big)}\Big)\vect m_i ,\quad \eta_i=n_i/n_{i+1} ,\quad \zeta_i=-\vect m_i\cdot\vect d_i ,
  \label{eq:snell_vec}
\end{equation}
in which $n_i$ and $n_{i+1}$ are the refractive indices on the incident and the transmitted sides. If the expression under the square root in Eq.~(\ref{eq:snell_vec}) is negative, the ray is totally reflected. Those rays that intersect the same wall interface again before reaching the measurement plane are also excluded from the mapping. The rays are sorted into three categories. The rays that pass through all interfaces and reach the light sheet inside the pipe are referred to as valid rays. The reflected rays do not reach the symmetry plane, and the background rays reach the symmetry plane outside of the pipe. Therefore, we only use the valid rays to define the image-correction map. The boundaries between different ray categories are obtained with sub-pixel accuracy. The boundary between the valid rays and the reflected rays represents the image of the inner interface, and the boundary between the reflected rays and the background rays represents the image of the outer interface. These boundaries provide the outlines of the pipe walls in the image space. They are used to fit the ray-tracing model to the wall contours detected in the recorded images (section~\ref{sec:registration}) and to determine the distance of each point to the wall.

\subsection{Wall detection}
\label{sec:registration}

\begin{figure}[tbp]
  \centering
  \includegraphics[width=\linewidth]{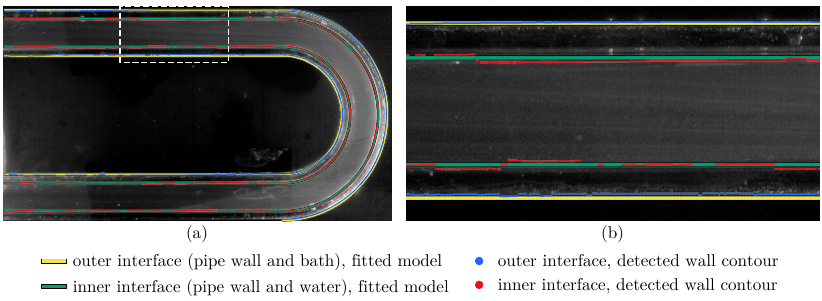}
  \caption{Predicted (outer yellow, inner green) and detected (outer blue, inner red) wall contours: (a) measurement domain and (b) zoom of the dashed box in (a).}
  \label{fig:wall_fit}
\end{figure}

To determine the two in-plane offsets of the camera, we fit the wall contours predicted by the ray-tracing model to the wall contours detected in an averaged image. Figure~\ref{fig:wall_fit}(a) shows these wall contours, and figure~\ref{fig:wall_fit}(b) shows an enlarged view of the dashed box in (a).

In the first step, we detect the contours in figure~\ref{fig:wall_fit}(a, b) from the brightness peak along the normals of the predicted boundaries within \num{12}\,px of the boundary. If no peak is found, the position of the steepest brightness change is used instead. The predicted contours of the inner and outer interfaces are represented by $N$ points, and for each point $k$ we denote the distance in pixels to the nearest detected contour point by $e_k$. False detections should not dominate the fit. Therefore, in the next step, we fit the two in-plane offsets by minimising the mean penalty
\begin{equation}
  J=\frac{1}{N}\sum_{k=1}^{N}H(e_k),
  \qquad
  H(e)=\begin{cases} e^2/(2e_0), & e\le e_0,\\ e-e_0/2, & e>e_0,\end{cases}
  \label{eq:registration}
\end{equation}
with $e_0=\num{3}$\,px. The penalty is of Huber type, with a quadratic part near the detected contour and a linear part beyond $e_0$. We perform the wall fit once on one averaged image. We then use the correction map computed from this fit for all measurements.

\subsection{Image correction}
\label{sec:correction}

\begin{figure}[tbp]
  \centering
  \begin{minipage}[b]{0.325\linewidth}
    \centering
    \includegraphics[width=\linewidth]{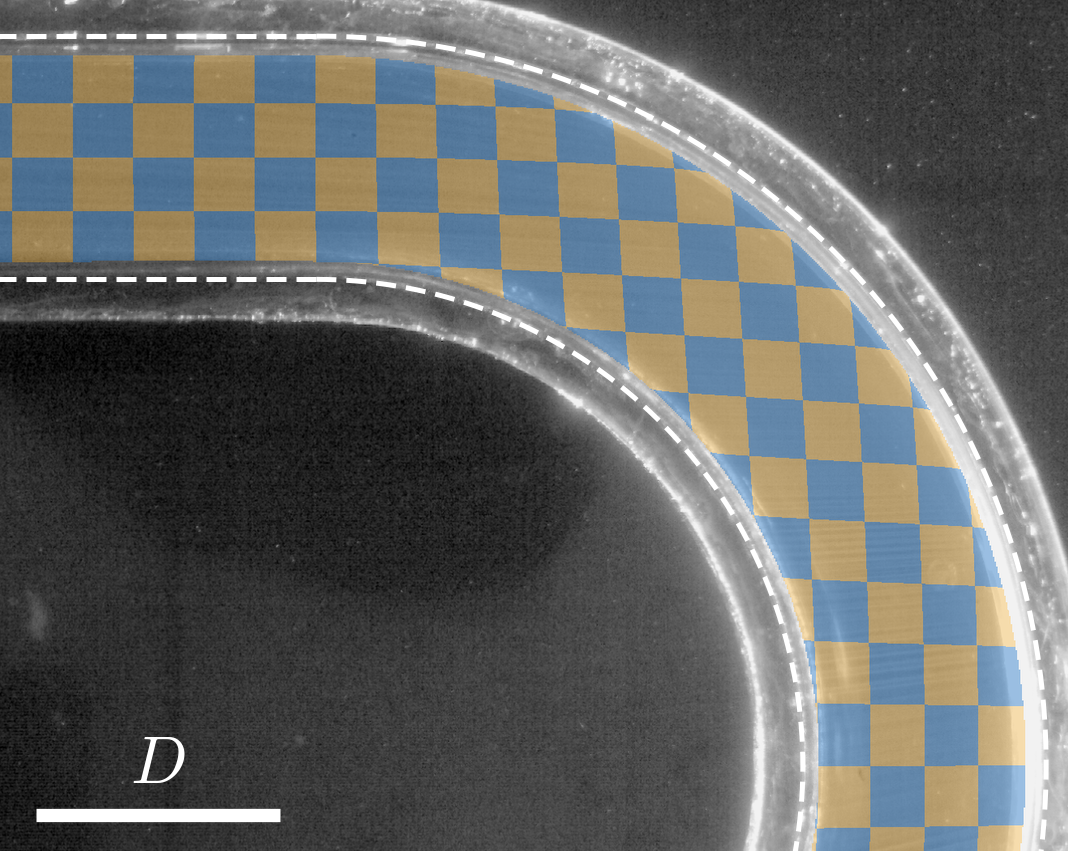}\\[1pt]
    {\small (a)}
  \end{minipage}\hfill
  \begin{minipage}[b]{0.325\linewidth}
    \centering
    \includegraphics[width=\linewidth]{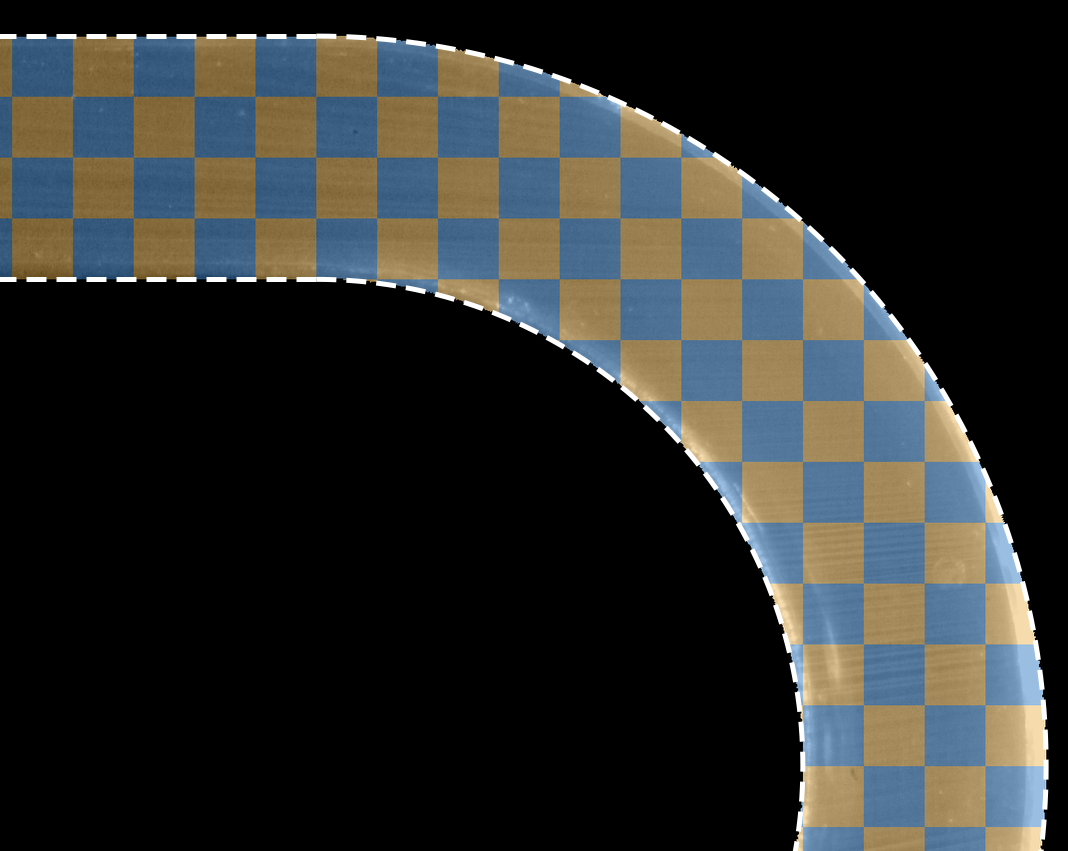}\\[1pt]
    {\small (b)}
  \end{minipage}\hfill
  \begin{minipage}[b]{0.325\linewidth}
    \centering
    \includegraphics[width=\linewidth]{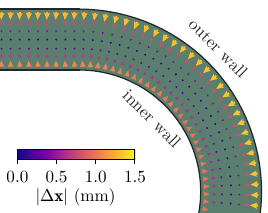}\\[1pt]
    {\small (c)}
  \end{minipage}
  \caption{Image correction in the first half of the bend: (a) raw and (b) corrected averaged image, and (c) displacement $\Delta\vect x$ from the true to the raw position of a point.}
  \label{fig:wall_detection}
\end{figure}

Using ray tracing, we find the following mapping from a sensor pixel $(i,j)$ to the symmetry plane at $\vect x$,
\begin{equation}
  \mathcal R:\ (i,j)\longmapsto\vect x.
  \label{eq:forward_map}
\end{equation}
$\mathcal R$ assigns to every pixel with a valid ray the point at which this ray intersects the symmetry plane.
For image correction, we in fact need the inverse mapping, from a point in the physical space to the pixel on which it is imaged,
\begin{equation}
  \mathcal M:\ \vect x\longmapsto(i,j),
  \label{eq:donor_map}
\end{equation}
in which $\mathcal M$ is found numerically by inverting $\mathcal R$ where $\mathcal M$ is single valued. However, in the complex and shadow zones of a refracting cylinder \citep{PaolilloAstarita2022Refraction}, $\mathcal M$ is multi-valued or empty. These zones are therefore excluded from the map instead of blending the grey values of several pixels in the complex zones as in \citet{Martins2018_Ray_Tracing_Based}. This is because blending would superpose multiple particle images from different points on the light sheet. As a consequence of this choice and using only the valid rays, the solid walls and the background are automatically removed from the corrected images. We build $\mathcal M$ from the pixels whose rays are categorised as valid. For a valid ray $m$, $(i_m,j_m)$ denotes its pixel and $\vect x_m=\mathcal R(i_m,j_m)$ is its intersection point with the symmetry plane. The triangulation of these intersection points divides the symmetry plane of measurement into triangles, and inside each triangle, the pixel coordinates are interpolated linearly,
\begin{equation}
  \mathcal M(\vect x)=\sum_{m=1}^{3}\lambda_m(\vect x)\,(i_m,j_m),
  \qquad\sum_{m=1}^{3}\lambda_m=1,
  \label{eq:donor_interp}
\end{equation}
in which the summation is over the three corners $\vect x_m$ of the triangle that encloses $\vect x$, and $\lambda_m$ are the barycentric coordinates of $\vect x$ in this triangle.
 Using the mapping $\mathcal M$, the corrected image can be obtained for the flow region as
\begin{equation}
  I_{\mathrm{corr}}(\vect x)=I_{\mathrm{raw}}\!\left(\mathcal M(\vect x)\right).
  \label{eq:remap}
\end{equation}
Here, $I_{\mathrm{raw}}$ is the grey value of the raw image as a function of the pixel position $(i,j)$, and $I_{\mathrm{corr}}$ is the corrected grey value as a function of the point $\vect x$ in the symmetry plane. The corrected image is obtained by evaluating Eq.~(\ref{eq:remap}) on a regular numerical grid of points in the symmetry plane, and $I_{\mathrm{raw}}$ is interpolated at the non-integer pixel position $\mathcal M(\vect x)$. Figure~\ref{fig:wall_detection}(a) shows the raw averaged image, and figure~\ref{fig:wall_detection}(b) shows the same image corrected using Eq.~(\ref{eq:remap}).

The walls and the background are automatically removed from the ray-tracing map, as seen in figure~\ref{fig:wall_detection}(b). The overlaid checkerboard is only meant to visualise the local image distortion, and its squares do not represent the regular grid onto which the images are resampled. The displacement shown in figure~\ref{fig:wall_detection}(c) is the distance between the true position of a point and the position at which it appears in the raw image. This displacement shows that the image of the flow is compressed towards the pipe axis. In the central part of the pipe, the compression is about $11\,\%$, while close to the walls the compression is stronger. Consequently, near the walls the raw image of a point appears $0.06$ to $0.08\Dpipe$ closer to the axis than the point itself.

The present method differs from earlier ray-tracing correction methods \citep{Zha2016Distortion,Martins2018_Ray_Tracing_Based,Akkurt2022Dimpled} in the combination of the following features. The optical interfaces are taken from the triangulated CAD surfaces, and within each triangle, the surface normals are interpolated. This allows for an accurate correction in the regions where the magnification varies within an interrogation window. As the camera model is aligned with the wall contours detected in the averaged image (section~\ref{sec:registration}), no calibration target is needed inside the pipe, and the model does not require extrapolation beyond a calibrated region \citep{PaolilloAstarita2022Refraction}. The alignment requires images from only one camera, in contrast to the self-calibration methods common in stereoscopic PIV \citep{Wieneke2005SelfCalibration}. The residuals of this alignment provide a measure to quantify the uncertainty of the vector positions and the ensuing uncertainty in the velocity measurements (section~\ref{sec:val_error}). Furthermore, the rays are categorised into different groups, and only rays reaching the laser sheet through the water are used for the mapping (section~\ref{sec:raymodel}). The correction approach excludes the regions where the mapping is multi-valued. Therefore, each corrected pixel is attributed to a single point on the laser sheet (section~\ref{sec:correction}). As every pixel needs a single reverse ray, the mapping from the camera sensor to the laser sheet is found without any iterative search and needs to be computed only once. Another consequence of this approach is that the solid walls and the image background are automatically removed from the corrected images.

\section{PIV acquisition and evaluation}
\label{sec:piv}

\subsection{Data acquisition}

The flow is seeded with silver-coated hollow glass spheres with a mean diameter of $15$\,\textmu m and density \SI{1.6}{g/cm^3}.
The maximum particle Stokes number is $1.9\cdot10^{-3} \ll 0.1$, well below the value at which particles still follow the large-scale turbulence well \citep{Samimy1991}. A Continuum Terra 527-20m (Amplitude Laser, Pessac, France) pulsed laser with a wavelength of \SI{527}{nm} and a maximum pulse rate of \SI{10}{kHz} is used to illuminate the particles with a light sheet through the pipe's symmetry plane. The images are recorded using a Photron Fastcam Nova R5-4K (Photron, Tokyo, Japan) high-speed camera with a maximum resolution of $4096 \times 2304$\,px$^2$ and a maximum frame rate of 1250\,fps at full resolution in combination with a Voigtländer 55mm f1.2 Nokton SLIIs lens (Voigtländer GmbH, Fürth, Germany).

In all experiments, the frame rate is set to \num{1700}\,fps at a reduced resolution of $3584\times1984$\,px on a camera sensor area of $23.3\times12.9$\,mm, resulting in a spatial image resolution of \num{15}\,px/mm. At all Reynolds numbers, every two consecutive frames are used as an image pair (consecutive pairs overlap by one frame) with a pulse separation of \SI{588.2}{\micro s}. The measurement consists of \num{12845} image pairs in each case, and the flow from $2\Dpipe$ upstream to $8\Dpipe$ downstream of the bend is shown in the field plots in the current article.

\subsection{Evaluation}
\label{sec:evaluation}

Before performing the PIV evaluations, we high-pass filter the corrected images, then compute the correlation in several stages using a multigrid scheme with starting windows of $96$\,px down to a final interrogation window of $12$\,px (\SI{0.80}{mm}, $0.04\Dpipe$). The pipe walls reflect the light sheet into a bright band of about $15$\,px (\SI{1}{mm}) along the pipe walls. As the window extends $6$\,px to each side of its centre, the band affects every window centred within $15+6=21$\,px, or $0.07\Dpipe$, of the wall. For this reason, we separate the measured regions into two parts. The region $|r|>0.43\Dpipe$ is referred to as the near-wall band, and the remaining region, $|r|\le0.43\Dpipe$, as the core region of the pipe throughout this work. In the mean velocity profiles, the near-wall band is shown with a shaded grey region and in the contour plots, with dashed lines.
\section{Reference simulations}
\label{sec:dns}

We use direct numerical simulations of the same \SI{180}{\degree} bend ($\curv=0.2$) at $\ReD=10\,000$ and $20\,000$ to provide validation data for the measurements at these Reynolds numbers. Each simulation integrates in time the incompressible Navier--Stokes equations with seventh-order polynomials using the spectral-element solver Nek5000 \citep{Fischer2008_Nek5000}. The computational domain has a straight inlet section of \num{16}$\Dpipe$ and a straight outlet section of \num{24}$\Dpipe$. The turbulent inflow is generated with a divergence-free synthetic-eddy method \citep{Hufnagel2018,Lupi2025}. Unlike a recycled inflow, this method is free of spurious periodicity \citep{Hufnagel2018,bagheri2026swirl}. The simulations are described in more detail in our previous work \citep{bagheri2026,bagheri2026swirl}. Table~\ref{tab:dns} lists the friction Reynolds number $Re_\tau=u_\tau\Dpipe/(2\nu)$, in which $u_\tau$ denotes the friction velocity. The resolution in wall units is given along the streamwise, radial and azimuthal directions, $\Delta_s^+$, $\Delta_r^+$ and $\Delta_\theta^+$. The time-averaged velocity field of each simulation is sampled on the symmetry plane to compare with the measurements.

\begin{table*}[tbp]
  \caption{Friction Reynolds number, grid spacing in wall units in the straight inlet section, and averaging time of the simulations, with the directions of the spacings in the cross-section $A$ (right).}
  \label{tab:dns}
  \begin{minipage}[c]{0.68\textwidth}
  \begin{tabular*}{\linewidth}{@{\extracolsep\fill}cccccc}
    \toprule
    $\ReD$ & $Re_\tau$ & $\Delta_s^+$ & $\Delta_r^+$ & $\Delta_\theta^+$ & $2t_{\mathrm{avg}}u_\tau/\Dpipe$ \\
    \midrule
    10\,000 & 316 & 2.3--11.1 & 0.41--4.7 & 1.0--6.1 & 98 \\
    20\,000 & 575 & 3.2--10.6 & 0.22--3.6 & 0.92--5.6 & 25 \\
    \botrule
  \end{tabular*}
  \end{minipage}\hfill
  \begin{minipage}[c]{0.28\textwidth}
  \centering
  \includegraphics{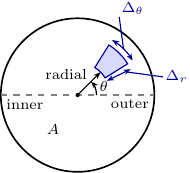}
  \end{minipage}
\end{table*}

\section{Validation}
\label{sec:validation}

\begin{figure*}[tbp]
  \centering
  \includegraphics[width=\linewidth]{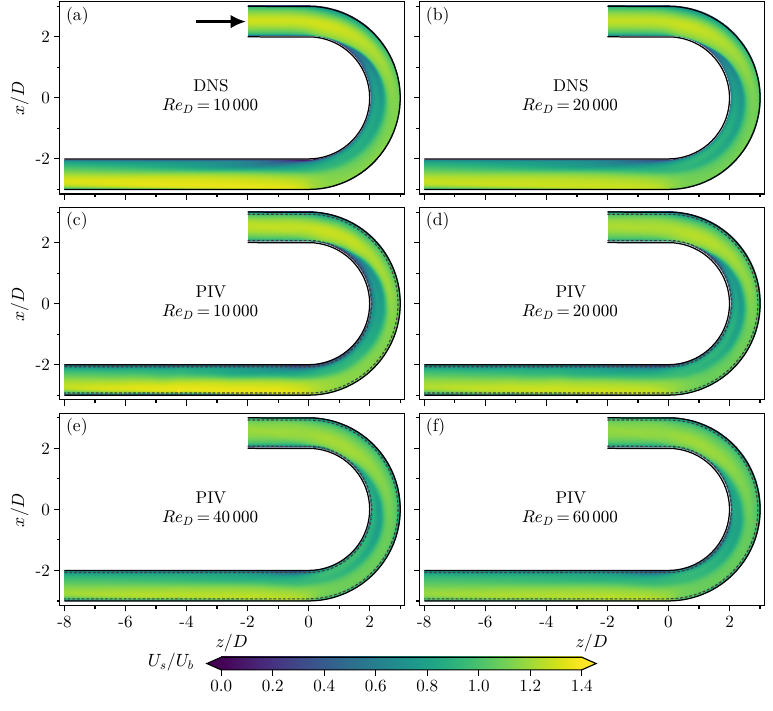}
  \caption{Mean streamwise velocity $\Us/\Ub$ in the symmetry plane. Note that the DNS domain is larger than what is shown in the panels (a) and (b). The dashed line shows the near-wall band $|r|>0.43\Dpipe$.}
  \label{fig:contours_us}
\end{figure*}

Figure~\ref{fig:contours_us} shows the mean streamwise velocities in the symmetry plane from the DNS and the PIV measurements. An initial comparison shows that both DNS and PIV results share the same flow features. Inside the bend, the flow accelerates near the outer wall and decelerates near the inner wall. Under a favourable pressure gradient on the outer wall at the bend exit, a fast fluid region extends up to the outlet section. With increasing Reynolds number, the flow becomes more uniform, and the difference between adjacent fast and slower regions reduces. A detailed validation of the PIV measurements (figure~\ref{fig:contours_us}(c, d)) against the DNS (figure~\ref{fig:contours_us}(a, b)) at $\ReD=10\,000$ and $20\,000$ is presented below. For the highest Reynolds number of $\ReD=60\,000$, we compare our results to the hot-wire data of \citet{Sudo2000}.

\subsection{Comparison with the simulations at $\ReD=10\,000$ and $20\,000$}
\label{sec:val_dns}

\begin{figure*}[tbp]
  \centering
  \includegraphics[width=\linewidth]{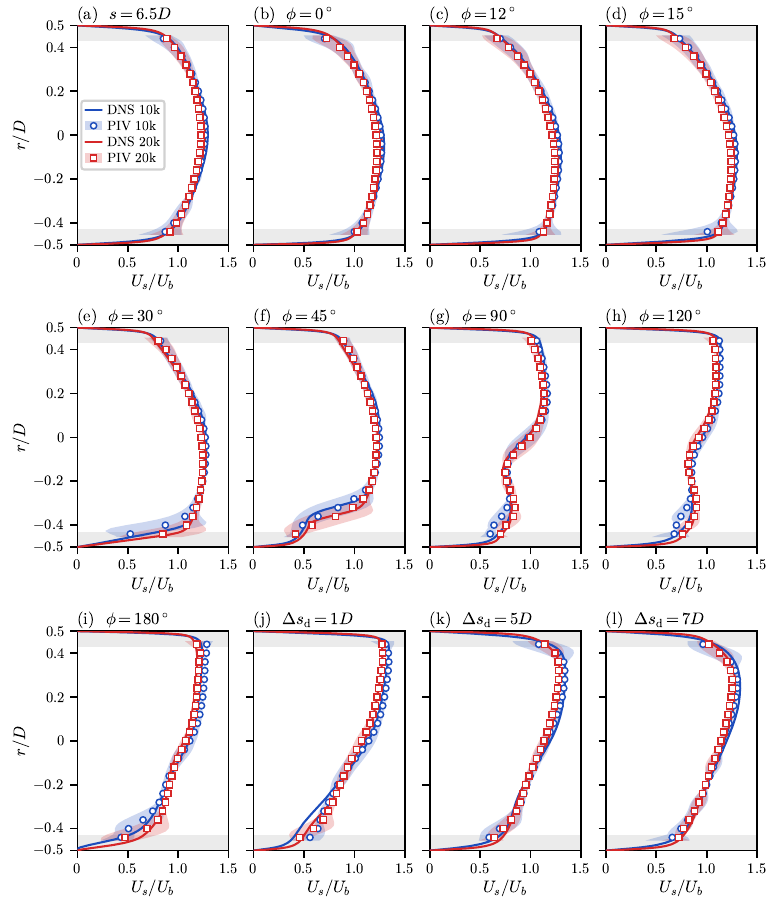}
  \caption{Mean velocity profiles $\Us/\Ub$ from the measurements (symbols) and the simulations (lines) at $\ReD=10\,000$ and $20\,000$.
  The shaded bands show the $95\,\%$ confidence interval of the measurement, Eq.~(\ref{eq:ci}), which combines the time-averaging and the position uncertainty modelled in section~\ref{sec:val_error}. The shaded grey region shows the near-wall band $|r|>0.43\Dpipe$.}
  \label{fig:profiles_val}
\end{figure*}

Figure~\ref{fig:profiles_val}(a--l) compares the mean velocity profile obtained from the measurement with the DNS results at twelve representative positions. Upstream of the bend (figure~\ref{fig:profiles_val}(a)), the root mean square of the difference between the measured and the simulated mean velocity profiles over the core region is $2.3\,\%$ of $\Ub$ at $\ReD=10\,000$ and $2.5\,\%$ at $20\,000$. Inside the bend, the velocity maximum moves towards the outer wall, and the low-momentum region forms near the inner wall in both the measurements and the simulations (figure~\ref{fig:profiles_val}, $\phi=\SI{45}{\degree}$ to $\SI{120}{\degree}$). As the Reynolds number increases, the velocity near the inner wall at the bend exit rises (figure~\ref{fig:profiles_val}, $\phi=\SI{180}{\degree}$), whereas the maximum of the mean velocity profile one diameter downstream of the bend exit decreases (figure~\ref{fig:profiles_val}, $\dsd=1\Dpipe$).

\subsection{Comparison with hot-wire data at $\ReD=60\,000$}
\label{sec:val_sudo}

\begin{figure*}[tbp]
  \centering
  \includegraphics[width=\linewidth]{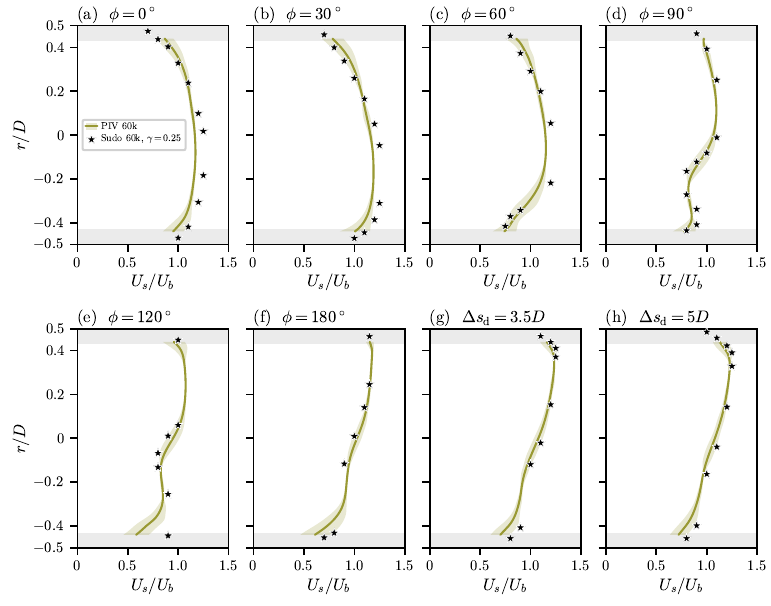}
  \caption{Streamwise velocity on the symmetry plane from the measurement at $\ReD=60\,000$ (line) and from the hot-wire data of \citet{Sudo2000} (stars). The shaded olive band shows the $95\,\%$ confidence interval of the measurement, Eq.~(\ref{eq:ci}), which combines the time-averaging and the position uncertainty modelled in section~\ref{sec:val_error}. The shaded grey region shows the near-wall band $|r|>0.43\Dpipe$.}
  \label{fig:sudo}
\end{figure*}

The comparison in figure~\ref{fig:sudo} is performed between the highest-Reynolds-number measurement, $\ReD=60\,000$, and the hot-wire data of \citet{Sudo2000}. Their curved pipe has a slightly stronger curvature ($0.25$ compared to $0.2$ in our study).

In the core region, the measured profiles follow the hot-wire data at all positions shown (figure~\ref{fig:sudo}(a--h)). The maxima of our profiles lie at $1.23$ to $1.24\Ub$ (figure~\ref{fig:sudo}, $\dsd=3.5\Dpipe$ and $5\Dpipe$), which is in close agreement with the $1.25$ from \citet{Sudo2000}. Over the inner half of the profile at lower bend angles, the hot-wire data are above our measurement by $0.07$ to $0.11\Ub$ (figure~\ref{fig:sudo}(a, b)). This is consistent with the stronger curvature in the hot-wire reference data. As curvature increases, the velocity maximum is displaced further towards the inner wall \citep{AlRafai1990}.

\subsection{Uncertainty quantification}
\label{sec:val_error}

For the cases where reference data are available, we quantify the agreement by the root mean square (rms) of the difference between the measured and the reference mean velocity profiles. Inside the core region, the measured profiles agree with the DNS within $4\,\%$ and $3\,\%$ of $\Ub$ at $\ReD=10\,000$ and $20\,000$, respectively, and with the hot-wire data within $6\,\%$ of $\Ub$ at $60\,000$. The largest differences lie near the edge of the near-wall band, where there is a strong velocity gradient.

In the absence of reference data, the measurement uncertainties need to be first modelled and then estimated. A common source of bias in PIV measurements arises from the spatial averaging of the velocity over the interrogation window with the width $W=0.04\Dpipe$ (equivalent to $12$\,px). This window deviation is the difference between the window average and the velocity at the centre of the window. We can estimate it by expanding the mean velocity at the window centre using a Taylor series to obtain
\begin{equation}
  \frac{1}{W}\int_{-W/2}^{W/2}\Us(r+\xi)\,\dd\xi-\Us(r)=\frac{W^2}{24}\,\frac{\dd^2\Us}{\dd r^2}+O(W^4).
  \label{eq:window}
\end{equation}
As the window width is a known parameter, the window deviation is a deterministic estimate. As long as the window lies inside the flow, the first-order term cancels due to the symmetry of the window about its centre. Therefore, the window deviation is a profile curvature contribution. This contribution is largest where the profile is strongly curved, namely along the inner-wall shear layer. Close to the wall, the window partly covers the wall \citep{Kaehler2012NearWall}. Its fluid part is no longer symmetric, and the first-order term remains. At all Reynolds numbers, we model the window deviation by Eq.~(\ref{eq:window}) with the curvature of the measured mean velocity profiles.

Apart from the inherent window deviation of PIV, the measurement results are subject to the following uncertainties. The first uncertainty arises from estimating the mean by finite-time averaging of the instantaneous velocity fields. The second source of uncertainty is in the position of the vectors, which follows from the alignment of the ray-tracing model with the wall contours in the ray-tracing step.

We model the time average as a normally distributed random variable. The time-averaging uncertainty depends on the standard deviation $\sigma$ of the streamwise velocity and on the number $N_{\mathrm{eff}}=t_{\mathrm{avg}}/(2T_I)$ of independent samples within the averaging time $t_{\mathrm{avg}}$. Here, $T_I$ denotes the integral time scale of this velocity, which we obtain at every node between the bend entry and $\dsd=7\Dpipe$ from the integral of the autocorrelation function up to its first zero crossing. By the central limit theorem, the normal model holds for any distribution of the velocity samples, as the time average is the mean of $N_{\mathrm{eff}}$ independent samples. The $95\,\%$ confidence interval of the time average is then \citep{SciacchitanoWieneke2016Propagation}
\begin{equation}
  \Us\pm1.96\,\frac{\sigma}{\sqrt{N_{\mathrm{eff}}}},
  \label{eq:ci_time}
\end{equation}
and its half-width is referred to as the time-averaging uncertainty.

\begin{figure}[tbp]
  \centering
  \includegraphics[width=\linewidth]{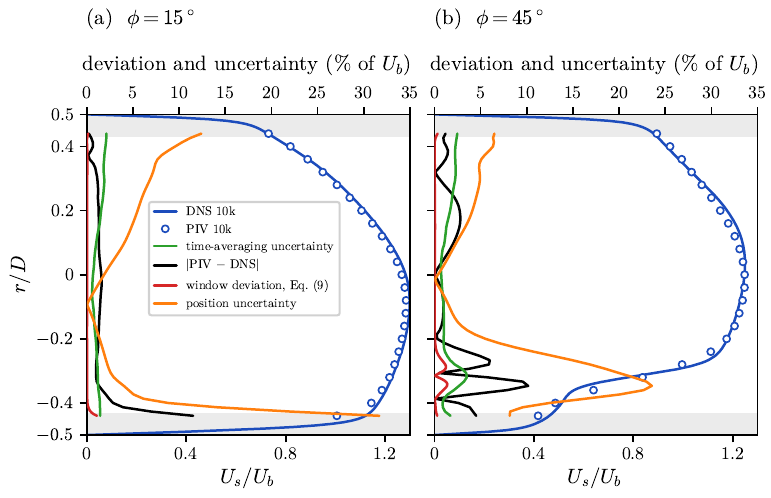}
  \caption{Mean velocity profiles of PIV and DNS (bottom axis), and the absolute difference between them, the window deviation, the position uncertainty and the time-averaging uncertainty (top axis) at $\ReD=10\,000$, (a) $\phi=\SI{15}{\degree}$, (b) $\phi=\SI{45}{\degree}$. The shaded grey region shows the near-wall band $|r|>0.43\Dpipe$.}
  \label{fig:error_decomp}
\end{figure}

The position uncertainty arises when the image correction places a vector at $r$ whose particles lie at $r+\delta$. As a result, the measured velocity differs from the true velocity at $r$ by
\begin{equation}
  \Us(r+\delta)-\Us(r)=\frac{\dd\Us}{\dd r}\,\delta+\frac{1}{2}\,\frac{\dd^2\Us}{\dd r^2}\,\delta^2+\cdots.
  \label{eq:position}
\end{equation}
The displacement $\delta$ of the vector positions inside the flow is unknown. Nevertheless, an estimate for $\delta$ at the walls is provided by $e_k$ in Eq.~(\ref{eq:registration}). Along the inner interface, the median $\tilde e$ of $e_k$ between the bend entry and $\dsd=7\Dpipe$ is $0.017\Dpipe$ (equivalent to $5.0$\,px).
The predicted wall contours follow from the fitted ray-tracing model, whereas each point of the detected wall contours in figure~\ref{fig:wall_fit} is found separately from the brightness profile of the averaged image. We assume the displacement between the fitted contour and the detected point to be a random variable $\delta$. This displacement contains the noise from the wall detection as well as any local deviation of the fitted ray-tracing model, and the two cannot be separated. We therefore attribute the whole displacement $\delta$ to the ray-tracing model. This provides an upper bound for the uncertainty of the vector positions and is a conservative estimate. We model $\delta$ by a normal distribution with zero mean and standard deviation $\sigma_\delta$. For a normal distribution, half of the values lie within $0.674$ standard deviations of the mean. Therefore, the standard deviation can be computed from the median of the residuals $\tilde e$ as
\begin{equation}
  \sigma_\delta=\frac{\tilde e}{0.674}=0.025\Dpipe.
  \label{eq:sigma_delta}
\end{equation}
Using this model, we can express the position uncertainty as a $95\,\%$ confidence interval. We keep the first-order term of Eq.~(\ref{eq:position}) and drop the higher-order terms. Then, $\delta$ lies within $\pm1.96\,\sigma_\delta$ with a probability of $95\,\%$, and the corresponding $95\,\%$ confidence interval of the mean velocity is
\begin{equation}
  \Us\pm1.96\,\sigma_\delta\left|\frac{\dd\Us}{\dd r}\right|,
  \label{eq:ci_position}
\end{equation}
and its half-width is referred to as the position uncertainty.
The position uncertainty is largest along the inner-wall shear layer and close to the walls, where the velocity gradient is largest.

As both the time-averaging uncertainty and the position uncertainty are modelled by independent normal distributions, the $95\,\%$ confidence interval of the measured mean velocity can be computed from
\begin{equation}
  \Us\pm1.96\sqrt{\frac{\sigma^2}{N_{\mathrm{eff}}}+\sigma_\delta^2\left(\frac{\dd\Us}{\dd r}\right)^2}.
  \label{eq:ci}
\end{equation}
The shaded bands in figures~\ref{fig:profiles_val} and~\ref{fig:sudo} show this interval. The window deviation is relatively small and is not part of the interval. It is given separately by Eq.~(\ref{eq:window}).

For figure~\ref{fig:error_decomp}, the window deviation is obtained by averaging the DNS mean velocity profile over the PIV window $W$. Figure~\ref{fig:error_decomp}(a, b) compares the window deviation, the position uncertainty and the time-averaging uncertainty with the actual difference between PIV and DNS at $\ReD=10\,000$. This shows that the difference between PIV and DNS inside the core region lies within the $95\,\%$ confidence interval of Eq.~(\ref{eq:ci}). Furthermore, most of this difference along the inner-wall shear layer is accounted for by the position uncertainty (figure~\ref{fig:error_decomp}(b)). In contrast, the window deviation grows only in the near-wall band.

Within the core region, the PIV measurements are validated against the DNS at $\ReD=10\,000$ and $20\,000$ and against the hot-wire data at $60\,000$. We use the identical setup and measurement method for $\ReD=30\,000$ to $50\,000$, therefore, the results for these intermediate Reynolds numbers are valid within the time-averaging and the position uncertainty of Eq.~(\ref{eq:ci}).
\FloatBarrier

\section{Analysis of PIV results}
\label{sec:results}

The turbulent flow in a \SI{180}{\degree} pipe bend shares a number of features that are affected by the Reynolds number and curvature. Most obviously, under the influence of the centrifugal force, the so-called Dean vortices form near the bend entry. Further downstream, a second vortex pair with the opposite direction of rotation to the Dean vortices appears. The appearance of these new vortices has been confirmed experimentally \citep{Rowe1970,Azzola1986,Anwer1989,Sudo2000,Lee2007} and was also reproduced in simulations \citep{Lai1991,Moulinec2005,bagheri2026swirl,Lupi2025}. However, the change of the onset of the second vortex pair with the Reynolds number has not been quantified. As a result of the secondary flow, the mean velocity profile becomes distorted, with its maximum near the outer wall and slow fluid near the inner wall \citep{Azzola1986,Sudo2000}. Downstream of \SI{90}{\degree} bends, the profiles were found to become more uniform at higher Reynolds numbers \citep{Hellstrom2013,Haroual2025}. However, the effect of Reynolds number on the mean turbulent velocity profile inside and downstream of \SI{180}{\degree} bends has not been quantified. Another feature of such flows is the presence of unsteady coherent structures inside and downstream of the bend. The effect of Reynolds number on the spatio-temporal features of the coherent structures, in particular on their frequencies, remains an open question \citep{KalpakliVester2015,Sundstrom2026}. In the following sections, we investigate the effect of Reynolds number on these three features between $\ReD=10\,000$ and $60\,000$. Using the radial velocity, we determine the onset of the second pair. The distortion of the mean streamwise velocity is investigated at various positions inside and downstream of the bend (section~\ref{sec:sec_flow}). Finally, the coherent structures are extracted and compared for the various Reynolds numbers against the DNS (section~\ref{sec:fhpod}).

\subsection{Effect of the Reynolds number on the mean flow}
\label{sec:sec_flow}

\begin{figure*}[tbp]
  \centering
  \includegraphics[width=0.98\linewidth]{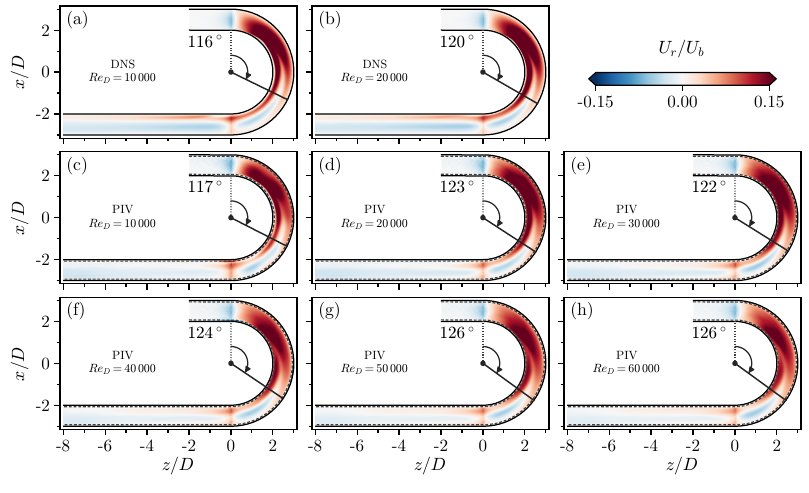}
  \caption{Mean radial velocity $\Ur/\Ub$ in the symmetry plane. The line from the bend centre and the arc mark the onset of the inward flow on the symmetry plane, the signature of the second vortex pair. Note that the DNS domain is larger than what is shown in the panels (a) and (b). The dashed line shows the near-wall band $|r|>0.43\Dpipe$.}
  \label{fig:contours_ur}
\end{figure*}

Figure~\ref{fig:contours_ur} shows the onset of the second vortex pair at different Reynolds numbers. The appearance of the second pair is found from the radial velocity on the symmetry plane, once it shows a zero crossing from positive to negative. Depending on the Reynolds number, the new vortex pair appears between $\phi=116^\circ$ and $126^\circ$. Upstream of the bend entry, the radial pressure gradient of the bend affects the incoming flow from the straight section and drives the fluid towards the inner wall. Downstream of the bend exit, the radial pressure gradient pushes the flow towards the outer wall. As a result, the radial velocity changes abruptly at the bend entry and at the bend exit under the influence of the radial pressure gradient and the absence of the centrifugal force to balance it.

\begin{figure*}[tbp]
  \centering
  \includegraphics[width=\linewidth]{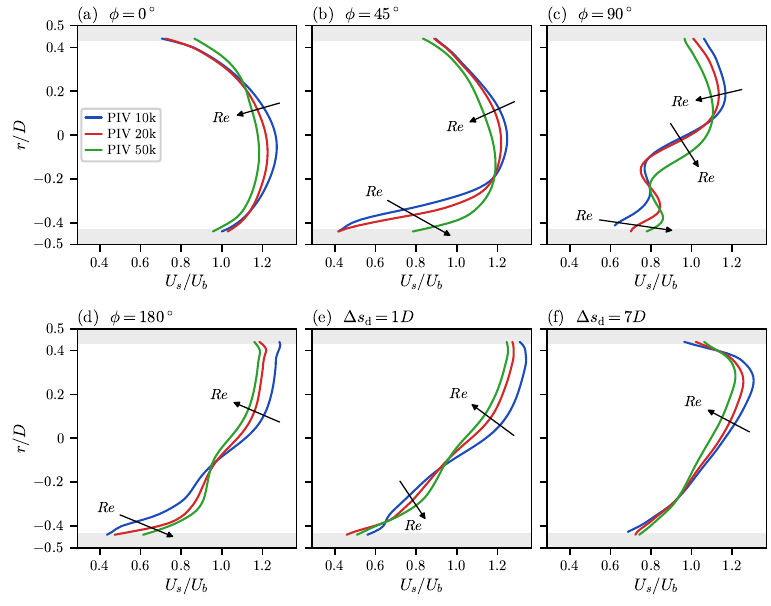}
  \caption{Mean velocity profiles $\Us/\Ub$ from PIV in the symmetry plane at $\ReD=10\,000$, $20\,000$ and $50\,000$, with arrows marking the change with increasing $\ReD$. The shaded grey region shows the near-wall band $|r|>0.43\Dpipe$.}
  \label{fig:profiles_re}
\end{figure*}

We compare the distortion of the mean flow at different Reynolds numbers. To illustrate the overall trend, figure~\ref{fig:profiles_re} shows three representative cases out of the six measured Reynolds numbers. As the Reynolds number increases, the measured mean velocity profiles become flatter in the bend and downstream of the bend exit (figure~\ref{fig:profiles_re}(c--f)). This means that the profile distortion becomes smaller as the Reynolds number increases.

The reduction of the distortion occurs because the incoming fully developed turbulent profile upstream of the bend becomes flatter. At $\phi=0^\circ$, the profile is flatter in the pipe centre and steeper near the walls at higher Reynolds number (figure~\ref{fig:profiles_re}(a)).

Furthermore, the events in the bend are shifted downstream when the Reynolds number increases. In the DNS, the onset of the inward flow of the second vortex pair is shifted from $\phi=116^\circ$ at $\ReD=10\,000$ to $120^\circ$ at $20\,000$ (figure~\ref{fig:contours_ur}(a, b)). In the measurements, it is shifted from $117^\circ$ at $10\,000$ to $126^\circ$ at $60\,000$ (figure~\ref{fig:contours_ur}(c--h)). Between $\ReD=50\,000$ and $60\,000$, the onset no longer changes and remains at $\phi=126^\circ$ (figure~\ref{fig:contours_ur}(g, h)). Near the inner wall from $\phi=45^\circ$ to the bend exit (figure~\ref{fig:profiles_re}(b--d)), the profile becomes steeper, and the layer of slow fluid becomes thinner as the Reynolds number increases. Downstream of the bend, the mean axial velocity profiles become more uniform, and the maximum near the outer wall decreases as the Reynolds number increases (figure~\ref{fig:profiles_re}(e, f)). The same behaviour was reported by \citet{Haroual2025} downstream of a \SI{90}{\degree} bend. Furthermore, the maximum of the mean axial velocity profile remains near the outer wall, consistent with \citet{Sudo2000} and \citet{Jain2022}.
\subsection{Coherent structures in the measured fluctuations}
\label{sec:fhpod}

Using DNS, \citet{bagheri2026swirl} identified four families of modes in the same configuration (a \SI{180}{\degree} bend with a curvature of $\curv=0.2$). However, the prohibitive computational cost of the DNS in \citet{bagheri2026swirl} did not allow higher Reynolds numbers to be explored. They therefore suggested examining whether the spatial structure of these modes and their Strouhal numbers persist at higher Reynolds numbers. Similarly, \citet{KalpakliVester2015} noted that the effect of Reynolds number on the frequency of swirl switching remains to be explored, and \citet{Sundstrom2026} called for a systematic mapping of this frequency as a function of the Reynolds number with measurements throughout the bend. The present PIV measurements address this question for the range of $\ReD=10\,000$ to $60\,000$.

In the analysis by \citet{bagheri2026swirl}, the mode with the lowest frequency corresponds to an axial wave, AX, at $\St\approx0.03$. The swirl-switching modes SS1 and SS2 and the swirl-breathing modes SB1 and SB2 form the second and third families and correspond to the sinuous and the varicose instability of the Dean vortices, respectively. The sinuous modes SS1 and SS2 displace the Dean vortices laterally. In contrast, in the varicose modes SB1 and SB2, both Dean vortices strengthen and weaken simultaneously. Both SS and SB families are confined to the curved section of the pipe and are unstable eigenmodes of the linearised Reynolds-averaged Navier--Stokes equations. Downstream of the bend, a shear layer forms between the second vortex pair and the Dean vortices. This results in the fourth family of modes (downstream shear-layer modes) abbreviated as DS1 and DS2. The detailed analysis of these four families of modes is given in \citet{bagheri2026swirl}.

On the symmetry plane where we perform the PIV, a two-dimensional representation of these three-dimensional structures can be found. The varicose SB modes are even with respect to the symmetry plane, and the in-plane velocity components remain finite on it. In contrast, the in-plane velocity components vanish for the SS, DS, and AX families due to their symmetry. However, these modes also have an even counterpart at the same frequency. This was shown for the SB mode, which is the even representation of the SS mode, both together capturing the instability of the Dean vortices. Therefore, the even representations of these modes can be captured using time-resolved PIV, and they are denoted AXe and DSe, with ``e'' standing for even.

To extract these structures from the instantaneous velocity fields, we use the filtered Hilbert proper orthogonal decomposition (FHPOD), as described by \citet{bagheri2026swirl}. In the first step, we remove the mean from the snapshots and band-pass filter them. The mean removal and the band-pass filtering follow the procedure of \citet{bagheri2026swirl} for this flow configuration at $\ReD=10\,000$. For a direct comparison of the planar PIV with the DNS, we repeat the same planar FHPOD on the DNS data in the symmetry plane. In the next step, each FHPOD mode of the PIV is phase-rotated until it is aligned in phase with the corresponding mode obtained from planar DNS snapshots at $\ReD=10\,000$.

\begin{figure*}[tbp]
  \centering
  \includegraphics[width=\textwidth]{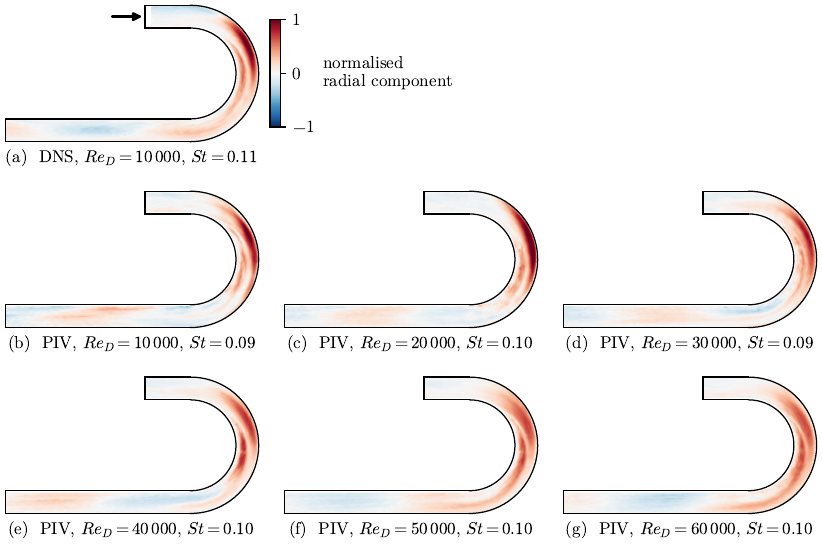}
  \caption{FHPOD mode SB1: radial component of (a) the DNS at $\ReD=10\,000$ and (b) to (g) the PIV at $\ReD=10\,000$ to $60\,000$.}
  \label{fig:fhpod_SB1}
\end{figure*}

\begin{figure*}[tbp]
  \centering
  \includegraphics[width=\textwidth]{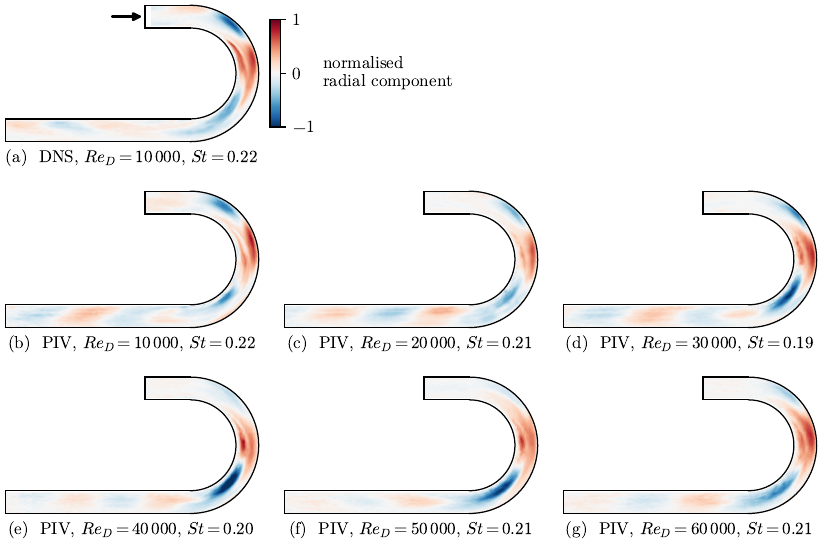}
  \caption{FHPOD mode SB2: radial component of (a) the DNS at $\ReD=10\,000$ and (b) to (g) the PIV at $\ReD=10\,000$ to $60\,000$.}
  \label{fig:fhpod_SB2}
\end{figure*}

The swirl-breathing modes are characterised by a cross-stream motion. We therefore show their radial velocity component, with the planar DNS in panel (a) and the PIV at $\ReD=10\,000$ to $60\,000$ in panels (b) to (g) of figures~\ref{fig:fhpod_SB1} and~\ref{fig:fhpod_SB2}. Both modes are strongest in the bend, where the instability of the Dean vortices originates. Their Strouhal numbers, $0.09$ to $0.10$ for SB1 and $0.19$ to $0.22$ for SB2, are close to those of the DNS, $0.11$ and $0.22$ (table~\ref{tab:fhpod}).

\begin{figure*}[tbp]
  \centering
  \includegraphics[width=\textwidth]{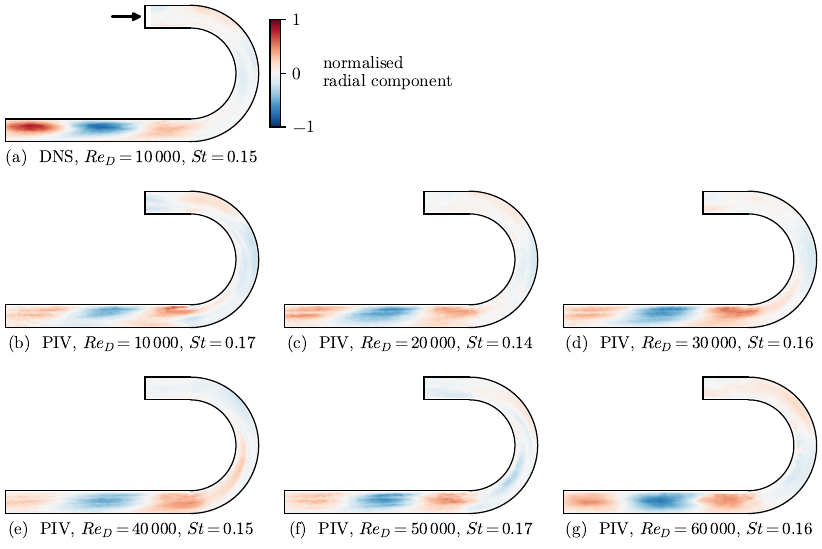}
  \caption{FHPOD mode DS1e, the even representation of DS1: radial component of (a) the DNS at $\ReD=10\,000$ and (b) to (g) the PIV at $\ReD=10\,000$ to $60\,000$.}
  \label{fig:fhpod_DS1e}
\end{figure*}

\begin{figure*}[tbp]
  \centering
  \includegraphics[width=\textwidth]{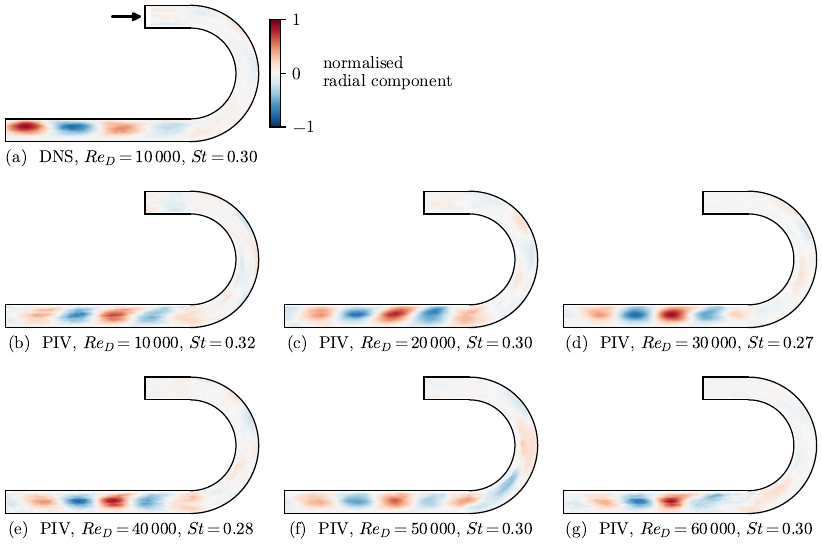}
  \caption{FHPOD mode DS2e, the even representation of DS2: radial component of (a) the DNS at $\ReD=10\,000$ and (b) to (g) the PIV at $\ReD=10\,000$ to $60\,000$.}
  \label{fig:fhpod_DS2e}
\end{figure*}

Downstream of the bend, the measurements capture the even representations DS1e and DS2e of the shear-layer modes. These modes are also shown by their radial velocity component, with the planar DNS in panel (a) and the PIV at $\ReD=10\,000$ to $60\,000$ in panels (b) to (g) of figures~\ref{fig:fhpod_DS1e} and~\ref{fig:fhpod_DS2e}. Both are waves in the outlet, where the shear layer between the second vortex pair and the Dean vortices forms downstream of the bend. The Strouhal numbers of DS1e and DS2e, $0.14$ to $0.17$ and $0.27$ to $0.32$, are close to those of the DNS, $0.15$ and $0.30$ (table~\ref{tab:fhpod}).
\begin{table}[tbp]
  \caption{Strouhal number of each mode in the planar DNS at $\ReD=10\,000$ and its range in the PIV at $\ReD=10\,000$ to $60\,000$, with the Strouhal-number band of the temporal filter of each mode in \citet{bagheri2026swirl}. The suffix ``e'' denotes an even mode. As odd modes are not captured by the PIV, they are marked by (--). For the sake of completeness, the Strouhal numbers of the odd modes are also reported here from \citet{bagheri2026swirl}.}
  \label{tab:fhpod}
  \begin{tabular*}{\linewidth}{@{\extracolsep\fill}lccc}
    \toprule
    Mode name & Filter band & DNS & PIV \\
    \midrule
    Axial wave, odd (AX) & $0.00$--$0.08$ & $0.03$ & -- \\
    Axial wave, even (AXe) & $0.00$--$0.08$ & $0.04$ & $0.02$--$0.04$ \\
    First swirl switching (SS1) & $0.07$--$0.14$ & $0.13$ & -- \\
    First swirl breathing (SB1) & $0.07$--$0.14$ & $0.11$ & $0.09$--$0.10$ \\
    First downstream shear layer, odd (DS1) & $0.12$--$0.18$ & $0.15$ & -- \\
    First downstream shear layer, even (DS1e) & $0.12$--$0.18$ & $0.15$ & $0.14$--$0.17$ \\
    Second swirl switching (SS2) & $0.20$--$0.30$ & $0.25$ & -- \\
    Second swirl breathing (SB2) & $0.20$--$0.30$ & $0.22$ & $0.19$--$0.22$ \\
    Second downstream shear layer, odd (DS2) & $0.18$--$1.00$ & $0.30$ & -- \\
    Second downstream shear layer, even (DS2e) & $0.18$--$1.00$ & $0.30$ & $0.27$--$0.32$ \\
    \botrule
  \end{tabular*}
\end{table}

\begin{figure*}[tbp]
  \centering
  \includegraphics[width=\textwidth]{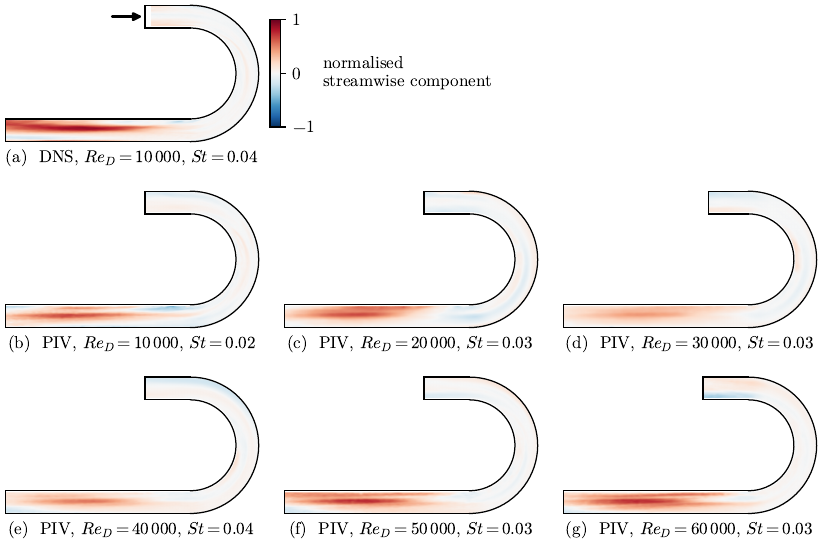}
  \caption{FHPOD mode AXe, the even representation of AX: streamwise component of (a) the DNS at $\ReD=10\,000$ and (b) to (g) the PIV at $\ReD=10\,000$ to $60\,000$.}
  \label{fig:fhpod_AX}
\end{figure*}

Finally, as AX is characterised by a streamwise wave, its streamwise velocity component is shown in figure~\ref{fig:fhpod_AX}. The figure shows the slow AXe structure in the outlet, where \citet{bagheri2026swirl} found its three-dimensional counterpart. In the measurements (figure~\ref{fig:fhpod_AX}(b) to (g)), as in the DNS (figure~\ref{fig:fhpod_AX}(a)), its streamwise component has one sign in the window. AX has the largest wavelength of all modes, about $22\Dpipe$ \citep{bagheri2026swirl}, which exceeds the $8\Dpipe$ outlet window. Nevertheless, the pipe continues downstream of the transparent part into the return line of the test bench (section~\ref{sec:setup}). Therefore, the mode can develop and is identified by the FHPOD in the measurements at all Reynolds numbers.
For all mode families, the Strouhal numbers are close to each other, while their scatter between the Reynolds numbers does not show a systematic trend, and they remain nearly constant from $\ReD=10\,000$ to $60\,000$. The spatial features of the modes are likewise invariant over this range of Reynolds numbers (figures~\ref{fig:fhpod_SB1} to~\ref{fig:fhpod_AX}). Considering that these modes are the nonlinear representation of the unstable local modes of the mean flow, and that the main features of the mean flow remain mostly unchanged (figures~\ref{fig:contours_us} and~\ref{fig:contours_ur}), this result is to be expected. This means that the coherent fluctuations keep their structure and their frequencies over this sixfold range of $\ReD$. The present measurements therefore answer this question for the families with a footprint in the symmetry plane.

\FloatBarrier
\section{Conclusions}
\label{sec:conclusions}

In this work, we perform PIV measurements for turbulent flow through a \SI{180}{\degree} pipe bend for $\ReD=10\,000$ to $60\,000$. The PIV measurements of such flows are particularly challenging. From a practical point of view, a full refractive-index matching requires changing the working fluid. These fluids are usually more viscous than the intended working fluid, and hence require higher pumping power at a given Reynolds number. Moreover, the refractive index of the fluid changes with temperature, and this problem becomes more severe due to the heat generated when a more viscous fluid is used. Therefore, an accurate measurement requires additional facilities to regulate the temperature of the working fluid during the measurement. The velocity profile in the bend and downstream of it is highly distorted. This creates regions with strong velocity gradients. As a result, the measurement is sensitive to small uncertainties in the position of the vectors that may arise from an imperfect refraction removal. The hybrid approach introduced here removes several of these difficulties. The refractions at the outer interface are removed by submerging the pipe inside an IOR-matched fluid bath, while water remains as the working fluid. The image distortion caused by the refractions at the inner interface is corrected using a reverse ray-tracing technique before calculating the velocities. The corrected images are then used to measure the particle displacements. The mean velocity profiles obtained using this hybrid approach are validated against DNS data at $\ReD=10\,000$ and $20\,000$ and prove to be quantitatively in agreement with DNS in the core region of the flow, more than $0.07\Dpipe$ away from the wall. Based on the rms of the difference between the measured and simulated mean velocity profiles, the agreement is within $4\,\%$ and $3\,\%$ of $\Ub$, respectively. The largest differences from the simulations are observed in the inner-wall shear layer of the bend and of the outlet section. The measurement at $\ReD=60\,000$ agrees well with the hot-wire data of \citet{Sudo2000}, obtained in a curved pipe with a slightly stronger curvature ($0.25$ compared to $0.2$ in our study), within $6\,\%$ of $\Ub$ in the same core region. For the intermediate Reynolds numbers $\ReD=30\,000$ to $50\,000$ where no reference data are available, we present a model for the uncertainty of the measured mean velocity profiles. As the setup and the measurement method remain identical, the velocity profiles are valid within the modelled uncertainty range in Eq.~(\ref{eq:ci}).

The turbulent flow through the \SI{180}{\degree} pipe bend shares the same main features at all investigated Reynolds numbers, although the local mean velocity profile varies with the Reynolds number. As the Reynolds number increases, the onset of the second vortex pair moves from $\phi=117^\circ$ at $\ReD=10\,000$ to $126^\circ$ at $60\,000$ in the measurements. Between $50\,000$ and $60\,000$, the onset remains at $126^\circ$. As the Reynolds number increases, the incoming fully developed turbulent profile becomes flatter, and the low-momentum region near the inner wall becomes less pronounced. As a result, the slow-fluid region close to the inner wall becomes thinner from $\phi=45^\circ$ to the bend exit. This results in a smaller distortion of the mean velocity profile in the bend compared to lower Reynolds numbers.

Furthermore, we apply the FHPOD to the time-resolved PIV data in the symmetry plane. We identify five distinct modes in the measurements, and their counterparts are found in the DNS. The mode shapes and the corresponding frequencies (Strouhal numbers) are in close agreement. Both the spatial features and the Strouhal numbers of these modes are found to be relatively insensitive to changes in the Reynolds number over the investigated range of Reynolds numbers.

The present study shows that PIV, together with the ray-tracing correction method, can be used for accurate time-resolved flow measurements in complex geometries. This is especially useful at high Reynolds numbers, where scale-resolving simulations are cost-prohibitive. The approach adopted here facilitates the measurement, as no special measures to change the working fluid and to handle the complexities associated with it are needed. Furthermore, no calibration target is required to be placed inside the geometry, where access may be difficult. As a result, the method has implications for the measurement of flows through complex geometries such as heat exchangers, junctions and valves. In addition to the mean velocity profiles, we show that spatio-temporal features of unsteady coherent structures can be extracted. The large-scale coherent structures are important from both a research and a practical perspective. The modes identified here are instabilities of the mean flow that contribute to sustaining the turbulence, and they induce pressure and wall shear stress fluctuations. As a result, the accurate identification of these structures has implications for cases where turbulence control or flow-induced noise and vibration may be of importance.

\backmatter

\bmhead{Authors' contributions}
E. Bagheri contributed to conceptualisation, methodology, data curation and software, performed the measurements, analysed the data, and wrote the manuscript. C. Heigl set up the optical system and the laser, performed the measurements with E. Bagheri, and contributed to the investigation and to writing the manuscript. F. Czwielong designed and built the test section and contributed to the investigation and to writing and reviewing the manuscript. S. Becker supervised the project, designed the test rig, reviewed the experimental work and the manuscript and provided feedback on both. P. Schlatter supervised the project, provided feedback, advised on the structure of the manuscript and contributed to its writing, review and editing.

\bmhead{Acknowledgements}
We thank Pia Trawiel and Judita Hudi of the Institute of Polymer Technology (Lehrstuhl f\"ur Kunststofftechnik, LKT) at Friedrich--Alexander--Universit\"at Erlangen--N\"urnberg for their help with the refractive-index measurements of the index-matching mixture.
The computing hardware was funded by the German Research Foundation (DFG). The authors gratefully acknowledge the scientific support and HPC resources provided by the Erlangen National High Performance Computing Centre (NHR@FAU) of the Friedrich--Alexander--Universität Erlangen--Nürnberg (FAU).

\section*{Declarations}

\bmhead{Funding}
This project was partially funded by the Bavarian Research Foundation (project AZ-1232-16). This work has also received funding from the European High Performance Computing Joint Undertaking (JU), together with Sweden, Germany, Spain, Greece and Denmark, under grant agreement no.~101093393.

\bmhead{Competing interests}
The authors have no conflicts to disclose.

\bmhead{Ethics approval and consent to participate}
Not applicable.

\bmhead{Consent for publication}
Not applicable.

\bmhead{Data availability}
The data will be provided upon request.

\bibliography{refs}

\end{document}